\documentclass{aa}
\usepackage{url}            
\usepackage{graphicx}
\usepackage[varg]{txfonts}
\usepackage{amsmath}
\usepackage{amssymb}            
\usepackage{amsfonts}           
\usepackage{color}              
\usepackage{siunitx}            
\newcommand{\be}{\begin{equation}}
\newcommand{\ee}{\end{equation}}
\newcommand{\bea}{\begin{eqnarray}}
\newcommand{\eea}{\end{eqnarray}}
\newcommand{\beq}{\begin{eqnarray}}
\newcommand{\eeq}{\end{eqnarray}}
\newcommand{\Alfven} {Alfv\'{e}n}
\newcommand{\RSUN}{$R_{\odot}$}
\newcommand{\GSUN}{$g_{\odot}$}
\newcommand{\DG}{\ensuremath{^\circ}}

\renewcommand{\bf}{}

\newcommand{\TotalEvents} {{47}}
\newcommand{\TypicalFibers} {{39}}
\newcommand{\TypicalFibersPos} {one}
\newcommand{\TypicalFibersNeg} {{38}}
\newcommand{\RopeFibers} {one}

\newcommand{\NarrowFibers}{three}
\newcommand{\NarrowFibersPos}{two}
\newcommand{\NarrowFibersNeg}{one}
\newcommand{\RopeFibersNeg} {one}
\newcommand{\FastDriftFibers} {four}
\newcommand{\FastDriftFibersPos} {three}
\newcommand{\FastDriftFibersNeg} {one}

\newcommand{\CC}[1]{\multicolumn{1}{c}{#1}} 

\graphicspath{ {./FiberFigures/}{./}}

\begin{document}

 \title{High resolution observations with Artemis--JLS}
 \subtitle{(II) Type IV associated intermediate drift bursts}
   \author{     C. Bouratzis \inst{1}
                        A. Hillaris \inst{1}
                        C.E. Alissandrakis \inst{2}
                        P. Preka-Papadema \inst{1}
                        X. Moussas \inst{1}
                        C. Caroubalos \inst{3}
                        P. Tsitsipis \inst{4}       \and
                        A. Kontogeorgos \inst{4} }
\offprints{C. Bouratzis}
\institute{                     Department of Physics, University of Athens, 15783 Athens, Greece\\ \email{kbouratz@phys.uoa.gr}
\and                            Department of Physics, University of Ioannina, 45110 Ioannina, Greece
\and                            Department of Informatics, University of Athens, 15783 Athens, Greece
\and                             Department of Electronics, Technological Educational Institute of Sterea Hellas, 35100 Lamia, Greece}
\authorrunning{Bouratzis et al.}
\titlerunning{Intermediate drift bursts}
 \date{Received .....; accepted ......}
\abstract
 {}
  {We examined the characteristics of isolated intermediate drift bursts and their morphologies on dynamic spectra, in particular the positioning of emission and absorption ridges. Furthermore we studied the repetition rate of fiber groups. These were compared with a model in order to determine the conditions under which the intermediate drift bursts appear and exhibit the above characteristics.}
{We analyzed sixteen metric type IV events with embedded intermediate drift bursts, observed with the Artemis--JLS  radio spectrograph from July 1999  to July 2005 plus an event on 1st August 2010. The events were recorded with the SAO high resolution (10 ms cadence) receiver  in the 270--450 MHz range with a frequency resolution of 1.4 MHz. We developed cross- and autocorrelation techniques to measure the duration, spectral width, and frequency drift of fiber bursts in \TotalEvents ~intermediate drift bursts (IMD) groups embedded within the continuum of the sixteen events mentioned above.  We also developed a semi-automatic algorithm to track fibers on dynamic spectra.}
{The mean duration of individual fiber bursts at fixed frequency was \mbox{ $\rm \delta t\approx300~ms$}, while the instantaneous relative bandwidth was { $\rm f_w/f\approx0.90\%$} and the total frequency extent was { \mbox{$\rm \Delta f_{tot}\approx 35~MHz$}}. The recorded intermediate drift bursts had  frequency drift, positive or negative, with average values of \mbox{$\rm df/fdt$} equal to  $\rm -0.027$ and 0.024\,s$^{-1}$ respectively. Quite often the fibers appeared in groups; the burst repetition rate within groups was, on average, \mbox{$\sim$0.98 s}. We distinguish six morphological groups of fibers, based on the relative position of the emission and absorption ridges. These included fibers with emission or absorption ridges only, fibers with the absorption ridge at lower or higher frequency than the emission, or with two absorption ridges above and below the emission. There were also some fibers for which two emission ridges were separated by an absorption ridge. Some additional complex groups within our data set were not easy to classify. A number of  borderline  cases of fibers with very high drift rate ($\rm \sim0.30~s^{-1}$) or very narrow total bandwidth ($\rm \sim8~MHz$) were recorded; among them there was a group of rope-like fibers characterized by fast repetition rate and relatively narrow total frequency extent. We found that the whistler hypothesis leads to reasonable magnetic field values ($\sim$4.6 G), while the Alfven-wave hypothesis requires much higher field. From the variation of the drift rate with time we estimated the ratio of the whistler to the cyclotron frequency, x, to be in the range of 0.3 to 0.6, varying by \mbox{$\sim$0.05--0.1} in individual fibers; the same analysis gives an average value of the frequency scale along the loop of $\sim$220\,Mm. Finally, we present empirical relations between fiber burst parameters and discuss their possible origin.}
{}
   \maketitle
%
\section{Introduction}\label{Intro}

The intermediate drift bursts (IDBs) represent a  class of fine structure embedded within the type--IV continuum \citep{Young61, Elgaroy73,Slottje1972,Fomichev78,Slottje1981}. They appear mostly in groups and their  principal characteristic is the drift rate, which is higher than the type--II but lower than the type III burst drift rate, hence their name. They include  typical fibers as well as   narrow--band rope--like fibers, first reported by \citet{Aurass1987b}~\citep[see also][]{Mann1989,Chernov1990,Chernov1990b,Chernov2006,Chernov2008};  there are other variants of these subclasses, such as the broadband fibers~\citep{Chernov2007} which were observed in the wake of type--II bursts, the fast drift bursts \citep{Jiricka01},  the narrowband fibers \citep{Chernov08B, nishimura2013, katoh2014},  and the long-duration slow-drift fibers \citep{Chernov2006}.
Apart from their  filament-like shape on the dynamic spectra \citep[reported as filaments with intermediate drift rate by][]{Chernov1990a}, another major characteristic of this type of fine-structure is their absorption-emission  ridge pair  form \citep[see reviews by~][]{Chernov2006, Chernov2011}. 

The intermediate drift  bursts are thought to represent the movement of a plasma wave exciter along a magnetic field  line in a coronal loop, usually post-flare.  Their radio emission is  interpreted as the result of whistler--Langmuir wave \citep{Kuijpers1972,Kuijpers1975,Kuijpers80} or \Alfven--Langmuir wave \citep{Bernold83} wave  interaction. \citet{Benz98}, on the other hand, proposed a combination of \Alfven~waves and electron cyclotron maser model for  intermediate drift bursts exciters propagating along overheated, low density magnetic structures which, possibly, correspond to  post flare loops \citep[see also reviews by ][where all the above mentioned mechanisms are compared]{Nindos07,Nindos08}. A  more recent approach \citep{Kuznetsov2006,Karlicky2013,zlobec2014} resorts to the modulation of the background type IV continuum by fast magneto-acoustic wave trains. According to the theory of whistler origin, the emission is enhanced at $\rm \omega_{pe} + \omega_w$ and reduced at $\rm \omega_{pe}$ where $\rm \omega_w$ is the whistler frequency. The frequency range of the emission is limited by the condition $\rm 0.25 \le \omega_w/ \omega_{ce} \le 0.5$; the first part of this relation is the necessary condition for the development of the instability and the second for strong cyclotron dumping. Due to their origin, the intermediate drift bursts qualify as coronal magnetic field diagnostics \citep[see~][]{Aurass2005,Rausche07,Rausche08}.

In a previous work \citep{Bouratzis2015} we  find that the  temporal association between energy release episodes and intermediate drift bursts is not as good as in other types of fine structure, such as pulsating structures and spikes. In that work we use the time of first impulsive energy release, evidenced from the first HXR or microwave peak. Intermediate drift bursts tend to appear well after the first energy release and the SXR peak; their histogram of time delay had two peaks at 10 and 40 minutes after the first impulsive energy release \citep[see Fig. 7 and 9 in~][]{Bouratzis2015}.  A similar behavior, albeit associating intermediate bursts with the decay phase of the hard X-ray flares, has been reported by \citet{Benz98}.

In the first article in this series \citep[][{hereafter referred to as paper I}]{Bouratzis2016} we examined the properties of narrowband bursts (such as spikes, etc.). In this article we present a detailed study of intermediate drift bursts observed during type IV events recorded by the ARTEMIS--IV\footnote{Now renamed \emph{ARTEMIS--Jean Louis Steinberg} \citep[see JLS obituary by ][]{Lequeux2016}} solar radio-spectrograph from July 1999  to July 2005 plus an event on 1st August 2010, in the 270-450 MHz frequency range.  We study the morphological characteristics of isolated members of this type of fine structure as well as of groups.

{ The observations and their analysis are described in Sect. \ref{Obs} with results presented in Sects. \ref{Classification}~and~ \ref{SingleBurst}. In Sect. \ref{Model} we present a simple fiber model based on the propagation of a whistler wave packet along a coronal loop  and we deduce parameters such as the magnetic field, the frequency scale length and the magnetic scale length along the loop. In Sect. \ref{relobs} we present empirical relations between fiber parameters and suggest possible interpretations. Our summary and conclusions are presented in Sect. \ref{discussion}.}

\section{Observations and data analysis} \label{Obs}
\subsection{Instrumentation}\label{Inst}

We used Type IV dynamic spectra recorded by the high sensitivity multichannel acousto-optical analyzer (SAO) of the Artemis-JLS solar radio-spectrograph operating at Thermopylae \citep{Caroubalos01,Kontogeorgos06}. The spectra cover the \mbox{$\rm 270\textnormal{-}450\,MHz$} range in 128 channels of \mbox{1.4 MHz} bandwidth, with a time resolution of 10\,ms. The dynamic spectra of the medium sensitivity, broadband, sweep-frequency receiver, {\bf Artemis-JLS/ASG} 
 \mbox{($\rm 650\textnormal{-}20\,MHz$ at 100\,ms)} were used to extend the frequency range when necessary.

In assembling the dataset we included all the type IV meter-wavelength events recorded by both \mbox{Artemis--JLS/ASG} and SAO during 1999-2010. Sixteen metric Type IV events \citep[listed in Table 2 of~][]{Bouratzis2015}, with well observed intermediate drift bursts were selected for further analysis; these include  a group of of rope-like fibers, four instances of fast drift burst (FDB) and three cases of narrowband fibers. The associated flares were evenly distributed in heliographic longitude.

\subsection{Data analysis}\label{DataAnalysis}

Our dynamic spectra were subject to high pass filtering along the time axis in order to suppress the slowly varying Type IV background emission \citep[see][]{Bouratzis2016}; this filtering enhanced rapidly varying structures, mostly fibers and pulsations which often overlapped. The separation of fibers from the pulsations was achieved with a second high pass filtering, this time along the frequency axis \citep[see][for a more detailed description of this method]{Kontogeorgos, Kontogeorgos08}. In most of the figures in this article we have included both filtered and unfiltered spectra, to ascertain that absorption features are not artifacts.

Based on the Artemis--JLS/SAO dynamic spectra, we classified the fibers in six morphological groups using the relative position of emission and absorption ridges as a criterion. These groups are described in Sect. \ref{Classification}. Furthermore, we developed techniques for measuring the observational characteristics of individual fibers and fiber groups, which are described below. 

\subsubsection{Analysis of individual fiber bursts}\label{DataAnalysisIndiv}
On the 10 ms SAO dynamic spectra, we measured the total bandwidth, $\Delta f_{\rm tot}$, the total duration, $\Delta t_{\rm tot}$ and the drift rate, $df/dt$, of 540 intermediate drift bursts (454 typical fibers, 72 FDBs and 14 rope-like fibers); the identification of individual bursts was done by inspection.

\begin{figure}[!]
        \begin{center}
                \includegraphics[width=\hsize]{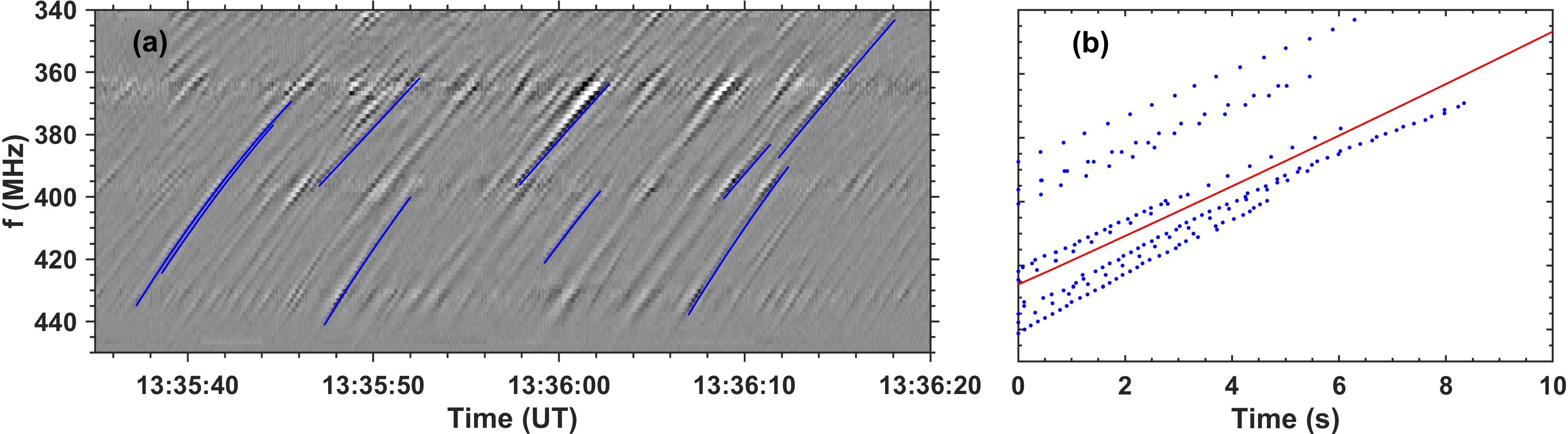}            
                \caption{Tracking of individual fibers detected in the same group as in Fig. \ref{CrossCorrVariation}, panel (a). The same tracks are presented as a function of time difference from the beginning in panel (b). The average track reconstructed from integration of the drift rate measured using cross-correlation is also plotted (red line). In this and subsequent figures the dynamic spectra are presented as negative images (darker parts corresponding to the higher intensities)}
                \label{Traced_CrossCorr}
        \end{center}
\end{figure}

Finally, we developed a semi-automatic  algorithm for the tracking of individual fibers on the dynamic spectrum. Once the beginning of a fiber was manually selected, the algorithm followed the peak of the flux from one frequency channel to the next. A similar method is also presented by Wang et al (2017). A total of 209 fibers were measured in this way.  Fig. \ref{Traced_CrossCorr} (a) shows the tracks of individual fibers detected within a group recorded on the 11 July 2011. In panel (b) are shown the tracks as a function of difference from start time. For comparison, we plot the average track reconstructed from the frequency drift (red line).

\subsubsection{Bulk parameters of fiber burst groups}\label{DataAnalysisBulk}
\begin{figure}[t]
\begin{center}
\includegraphics[width=\hsize, trim=1.4cm 0.0cm  0cm 0.0cm,clip]{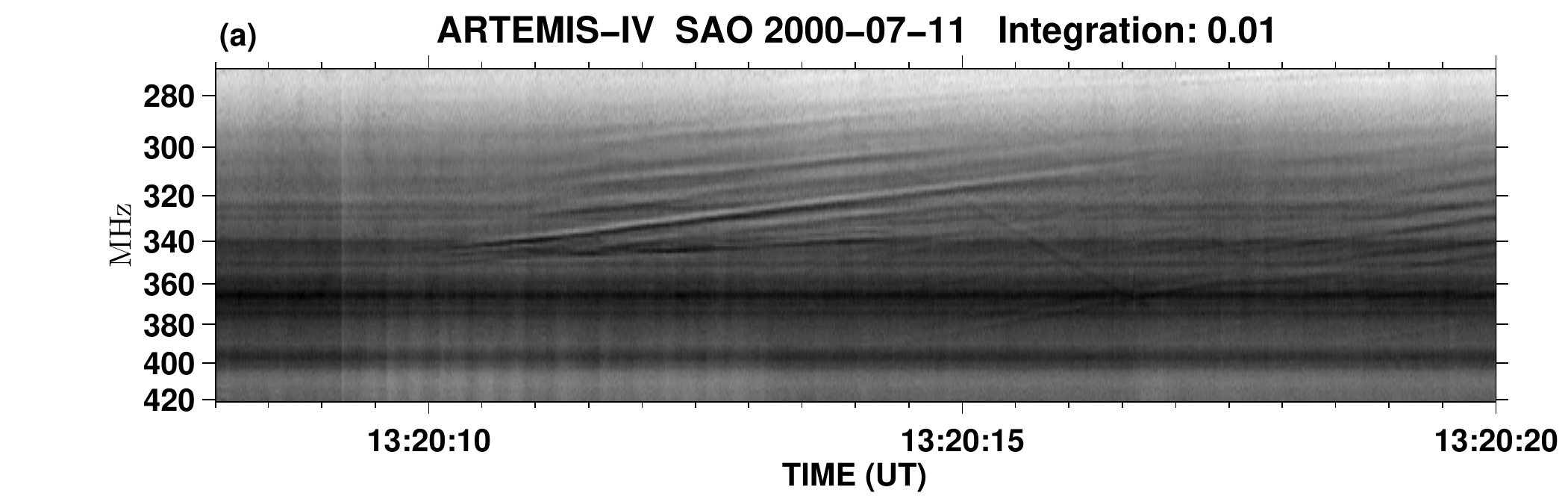}\\
\includegraphics[width=\hsize]{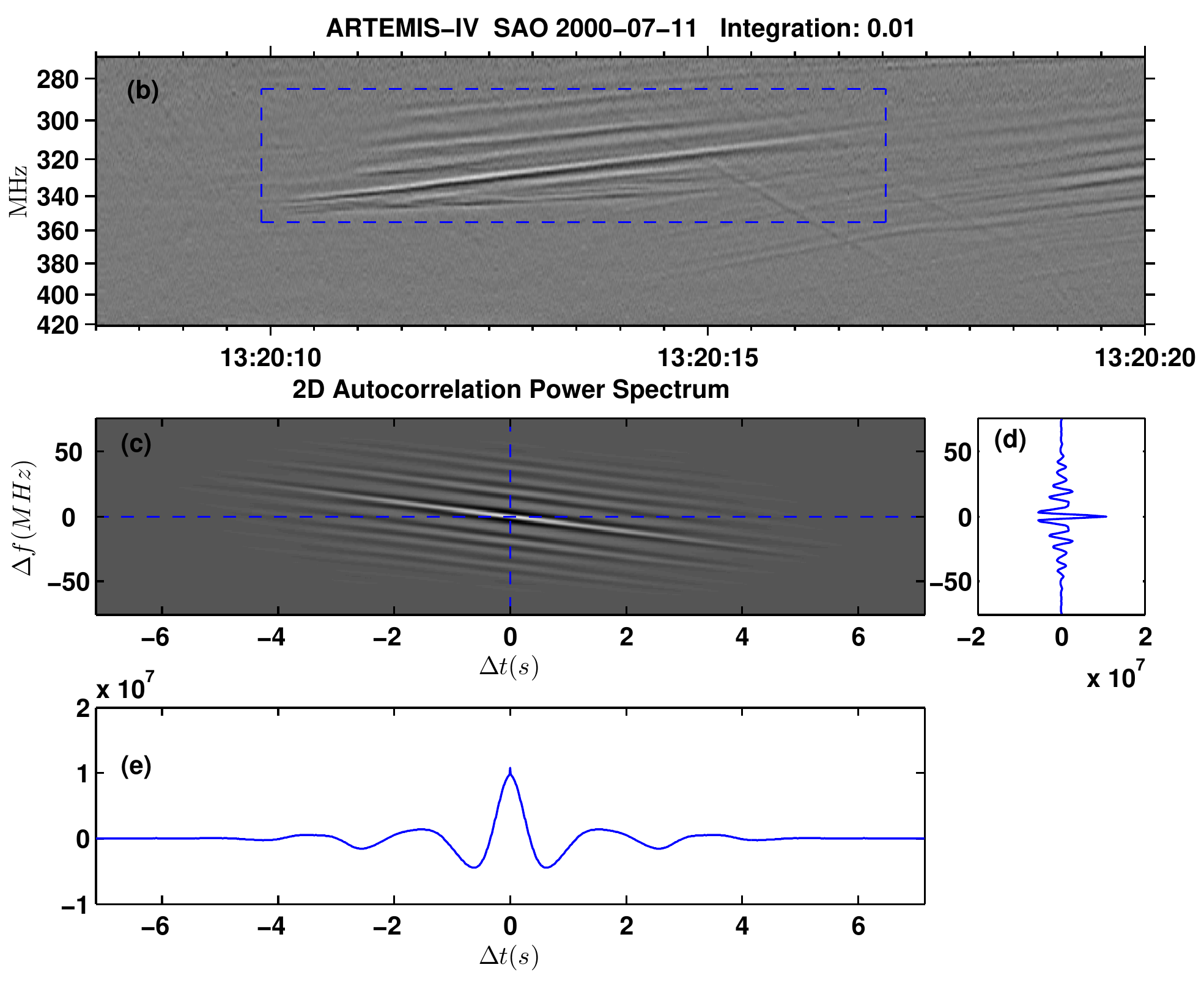}
\caption{(a): Original dynamic spectrum of a fiber group. (b) The same dynamic spectrum, filtered as described in Sect. \ref{DataAnalysis}. (c): The two-dimensional autocorrelation function of the region delimited by the box in (b); the time lag in seconds is on the horizontal axis. (d) \& (e): Profiles of the 2D autocorrelation function in the horizontal and the vertical direction, along the  blue dashed lines in panel (c). }
\label{2Dauto}
\end{center}
\end{figure}
We measured the bulk parameters of fiber groups by computing the 2D autocorrelation function of the dynamic spectrum of each group (Fig. \ref{2Dauto}). We note that the autocorrelation functions of the raw and filtered dynamic spectra are very similar although the spectra themselves look very different. The fiber duration, $\rm \delta t$, was computed as the full width at half maximum (FWHM) of the autocorrelation along the time axis, whereas the FWHM along the frequency axis gave the bandwidth, $\rm \delta f$, of the emission (Fig. \ref{2Dauto}d, e respectively).  The first maximum of the auto-correlation along the time axis gave the inverse of the repetition time, $T$, of individual fibers in the group. The inclination of the autocorrelation function gave the bulk frequency drift rate, $\rm \delta f / \delta t$, of the group.
Although this method gives average characteristics of the fiber bursts in a group rather than of individual bursts,  it is fast, accurate, and easy to implement.  The values of the parameters of interest differ less than  $\sim10$\% from those measured on individual bursts.

\begin{figure}[h]
\begin{center}
\begin{tabular}{ccc}    %
\includegraphics[width=0.5\textwidth]{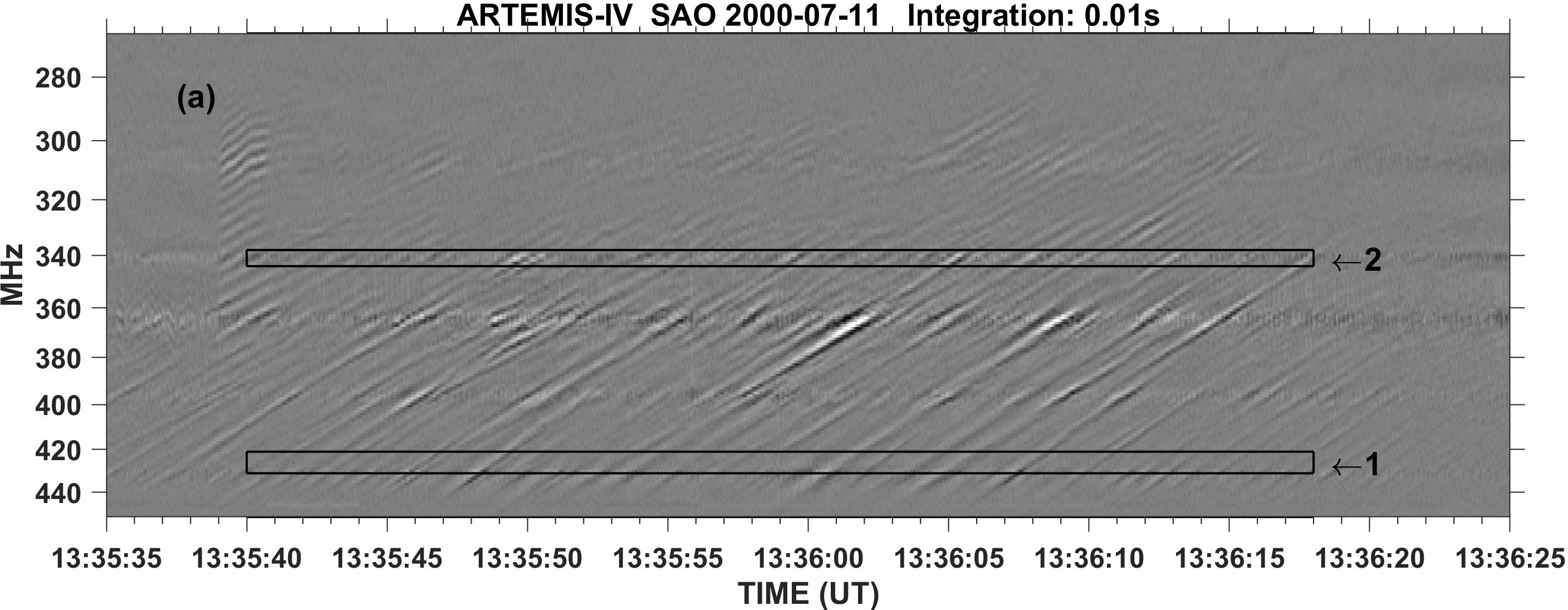}    \\
\includegraphics[width=0.5\textwidth]{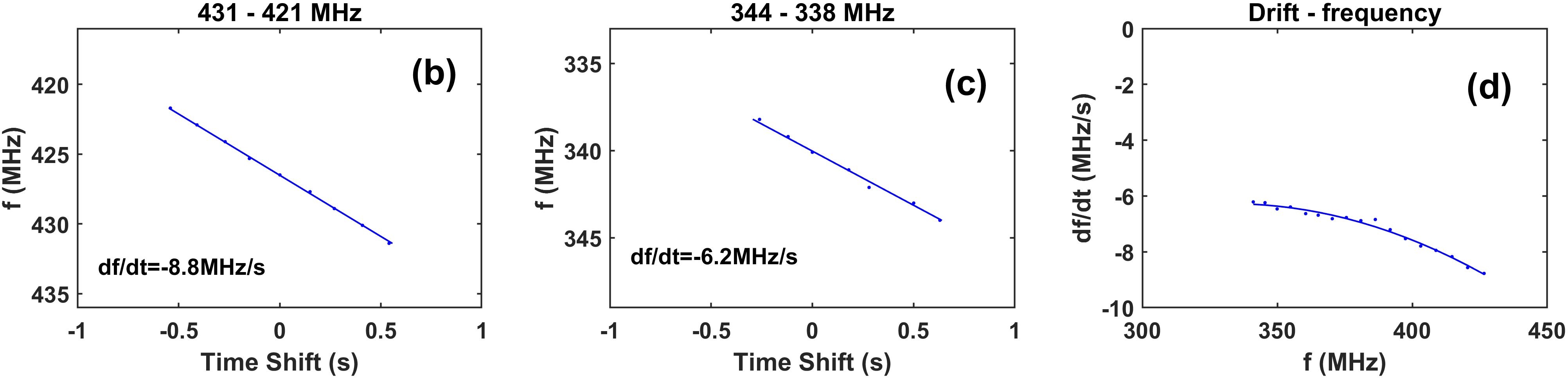}
\end{tabular}
\caption{ Dynamic spectrum from SAO (a) for the fiber group used in the computation of drift rate at a sliding  frequency window using cross-correlation on each position of the window. Panels (b) and (c) show the frequency as a function of time shift from the first and  last frequency window, labelled respectively with 1 and 2 in dynamic spectrum. Panel (d) shows the variation of the drift rate with frequency.}
\label{CrossCorrVariation}
\end{center}
\end{figure}

In addition to the average frequency drift rate of fiber groups that we computed with the 2D autocorrelation, the group drift rate as a function of frequency (and hence of time) was computed by selecting a ten channel spectral window and sliding it on the dynamic spectrum.  At each position of the sliding window the time shift of each channel with respect to the central one was computed by 1D cross-correlation, and from that the frequency drift was calculated.  An example  is presented  in Fig. \ref{CrossCorrVariation}. We note that the frequency drift rate decreases with time, which results in a characteristic curvature of the fiber burst in the dynamic spectrum.

\begin{figure}[!]
\begin{center}
\includegraphics[width=\hsize]{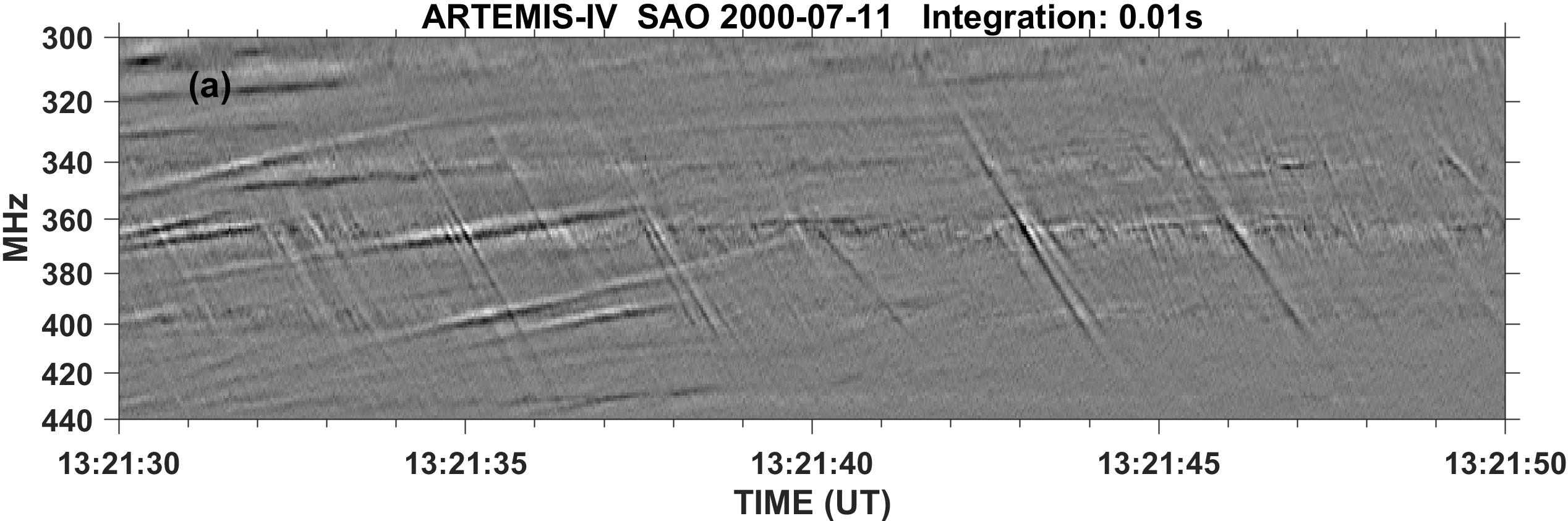}\\
\includegraphics[width=\hsize]{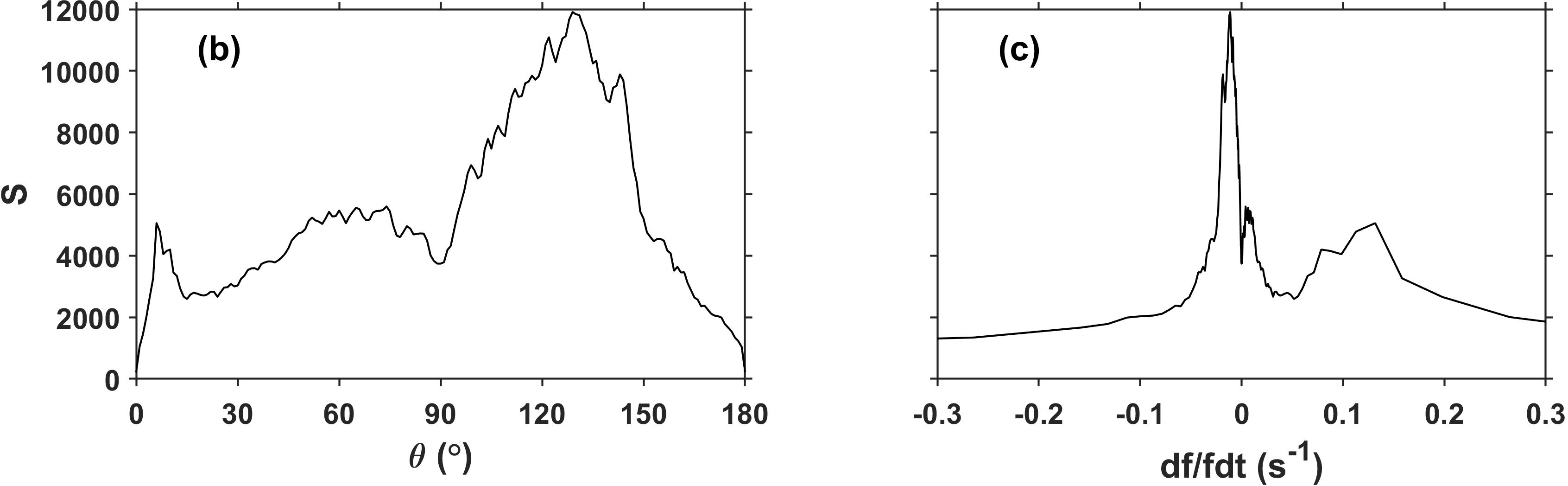}           
\caption{(a): Dynamic spectrum with two overlapping groups of intermediate drift bursts. (b):  Amplitude of the corresponding power spectrum as a function of the polar angle on the FT plane. (c): The same amplitude as a function of drift rate. The two peaks at 6\DG~and 129\DG~correspond to drift rates of 0.132 s$^{-1}$ (FDB) and \mbox{-0.012 s$^{-1}$} (typical fiber group) respectively.}
\label{radial}
\end{center}
\end{figure}

Quite often, two or more fiber groups with different drift rates overlap on the dynamic spectra. In such cases we computed the 2D Fourier transform of the dynamic spectrum, expressed it in polar coordinates and integrated the FT amplitude along the radial direction on the FT plane \citep{Tsitsipis06A,Tsitsipis07}. The spectral power, plotted as a function of the polar angle (Fig. \ref{radial}), helped us separate the fiber groups. Since the polar angle is associated to the inclination of the structures on the dynamic spectrum, the positions of the power spectra peaks provide the mean frequency drift rates.

A brief, comprehensive overview of the characteristics of each of the  \TotalEvents~intermediate drift bursts groups is presented in Table \ref{Dataset}. This dataset includes \TypicalFibers~groups of typical fibers,  \RopeFibers~of rope fibers, \FastDriftFibers~of fast drift fiber bursts (or fast drift bursts, in short FDB) and \NarrowFibers ~groups of narrowband fibers.

\section{Classification}\label{Classification}
In this section we describe our examination of the morphological classification of fiber bursts. We start with some general remarks (Sect. \ref{3W}), proceed with the typical fibers (Sect. \ref{TypicalFibers}) and then we treat the remaining  subcategories of the intermediate drift bursts (Sect. \ref{Ropes}, \ref{FDB}, \ref{NarrowFiber}, and \ref{ChainBurst}). 

\begin{figure}
\begin{center}
\includegraphics[width=\hsize]{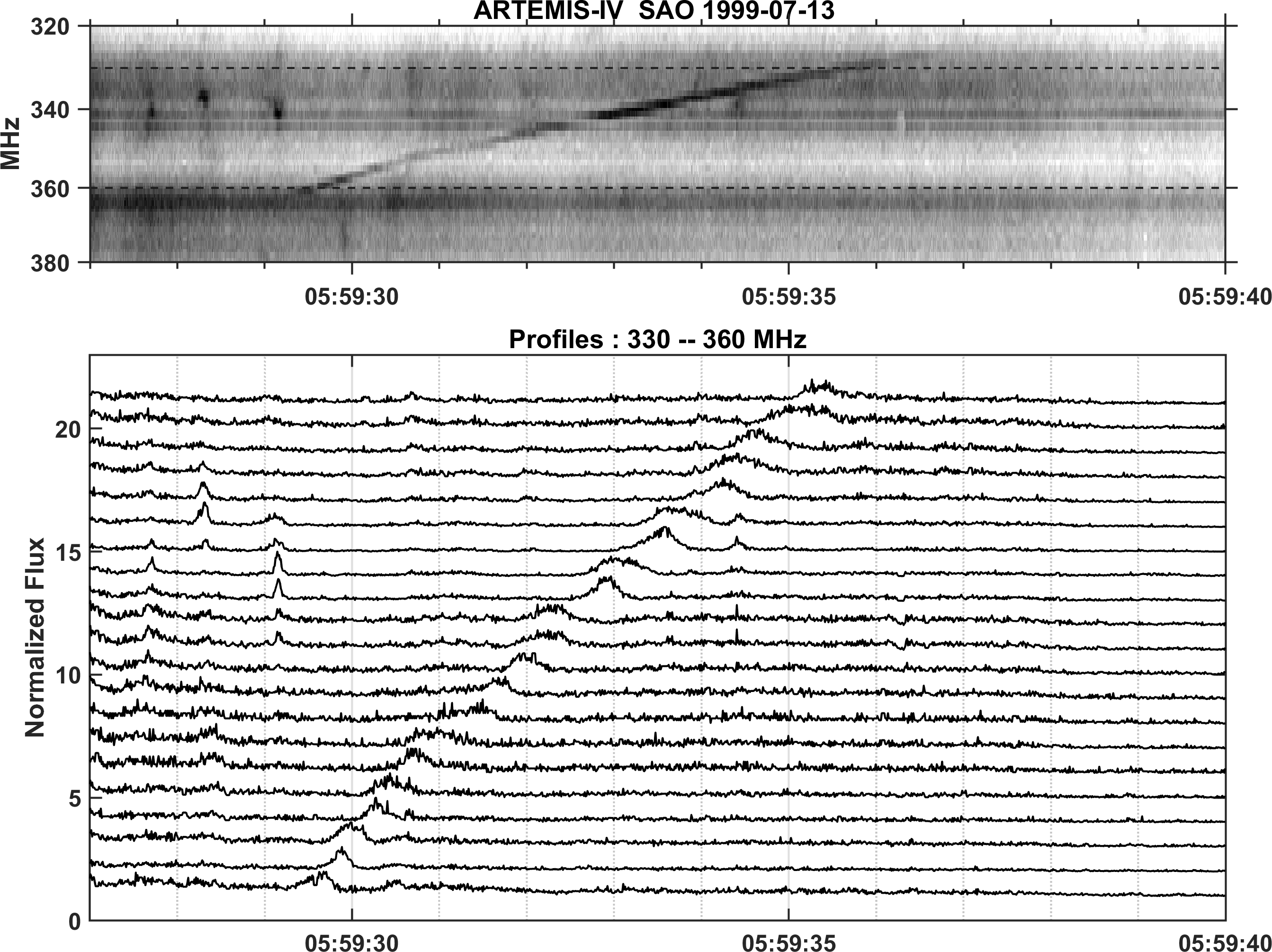} 
\caption{Isolated fiber with emission ridge only. Top: Filtered dynamic spectrum. Bottom: Intensity profiles}
\label{fiberIsoEmission01}
\end{center}
\end{figure}

\subsection{General remarks}\label{3W}
The distinctive attributes of the fiber bursts are the filament-like shape on the dynamic spectra and their absorption-emission  ridge pair  structure. As mentioned in Sect. \ref{Intro}, these characteristics have been interpreted in terms of a three-wave  coupling process ($\rm l+w\rightarrow t$), where whistlers (w) coalesce with the Langmuir waves (l) of the Type-IV continuum, producing the transverse wave emission (t) which escapes the solar corona \citep{Kuijpers1975}. The wave-wave interaction is thought to deplete the Langmuir wave energy, resulting in the absorption ridge at the low frequency side of the IDB. A $\rm l\rightarrow t+w$ three-wave decay \citep{Chernov1990a} on the other hand, may result in an absorption ridge at the high frequency side of the burst, as reported by \citet{Young61,Elgaroy73} and \citet{Jiricka1999}. The whistler-Langmuir wave coalescence is expected to produce {\bf an
emission ridge only,} without absorption when the type IV background is low due to suppression by means of induced scattering \citep{Kuijpers&Slottje1976}. When the emission is suppressed or scattered, the result may be fibers with pure absorption alone \citep{Elgaroy73}.

The magnetic geometry effects on the morphology of the intermediate bursts manifest themselves in the rope-like fiber bursts. These are chains of IDBs with negative frequency drift, relative  total frequency extent \mbox{$\Delta f \simeq 2\%$}, which is systematically smaller than the frequency extent of the typical  fibers. They also have pronounced absorption ridges at the low frequency side and repetition rate usually higher than the typical fibers.

The small frequency extent has been interpreted  \citep{Mann1989, Mann1990} as the result of whistler generation within localized magnetic structures which restrict the total bandwidth.  This implies that the magnetic field driving the exciter is not static, but is subject to MHD perturbations \citep{Mann1989}. Some examples of localized, magnetic MHD shock bound traps were also proposed for chains of ropes, drifting collectively at a rate comparable to type-II burst drift. These include, firstly, fast shock fronts in reconnection \citep{Chernov1990b, Chernov2006} and, secondly,   a type-II shock overcoming the leading edge of a CME \citep{Chernov2014}. The localized magnetic confinement, smaller than a loop, is thought to be responsible for the high repetition rate due to a higher bounce frequency of the trapped, loss cone electrons. The movement of this magnetic trap within the coronal plasma, on the other hand, is thought to be manifested as the chain frequency drift of the whole rope-like fiber group.

\subsection{Typical fibers}\label{TypicalFibers}
On the basis of the position of the emission and the absorption ridges, we distinguished six morphological groups of fibers:
\begin{figure}
\begin{center}
\includegraphics[width=\hsize]{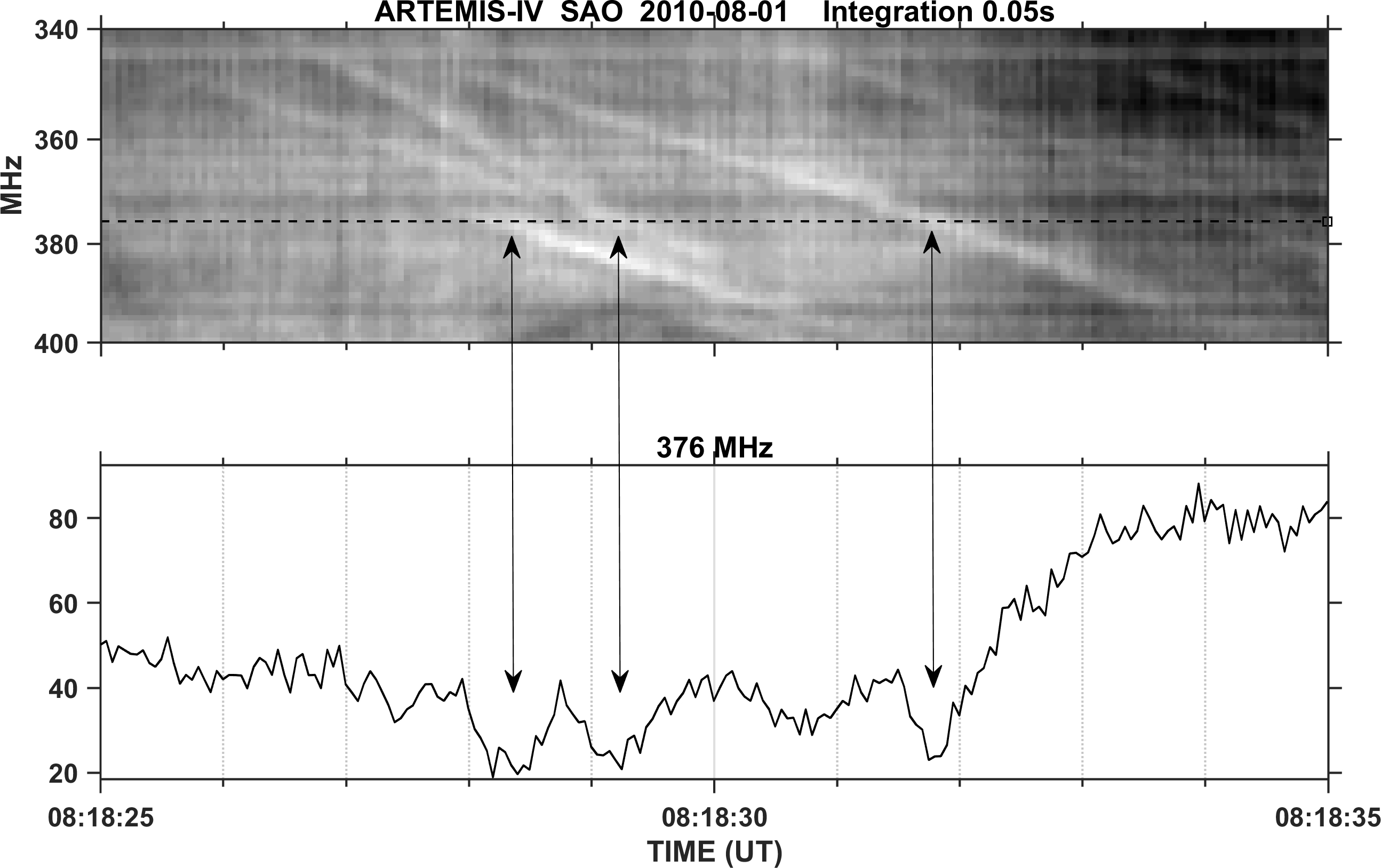}  
\includegraphics[width=\hsize]{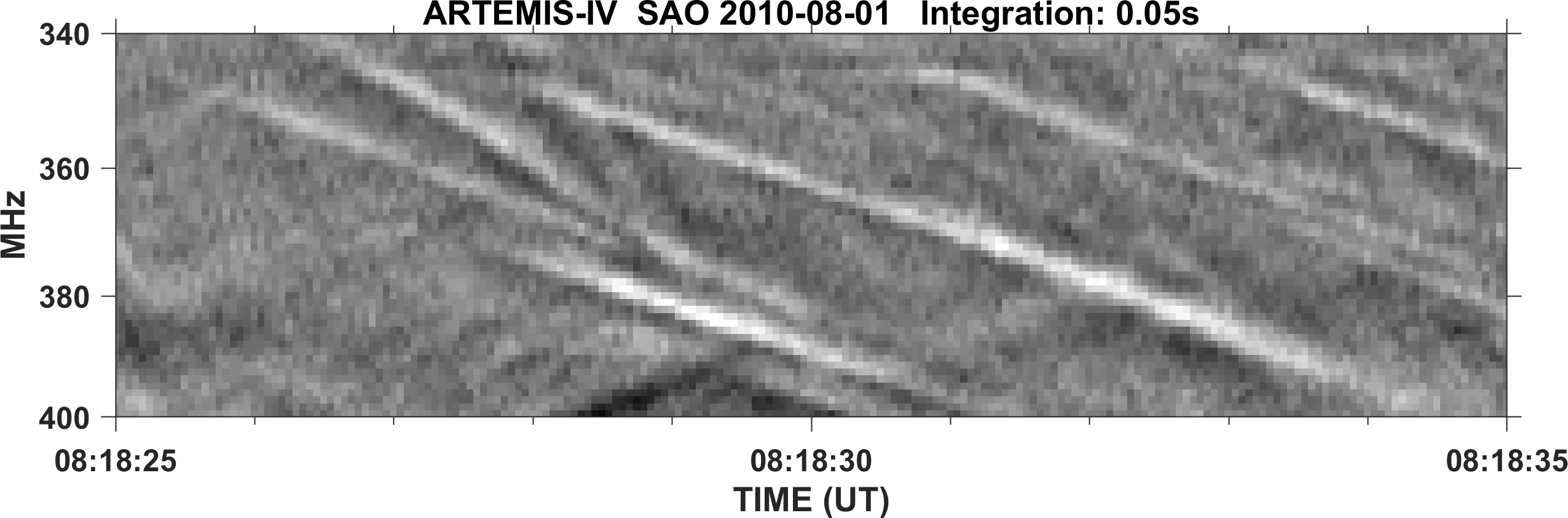}
\caption{Fibers appearing as absorption ridges on the Type IV continuum. Top:  Dynamic spectrum (raw, negative). Middle: Intensity profile at 376 MHz (dashed line in the top panel); arrows point to the absorption ridges. Bottom: Filtered dynamic spectrum.}
\label{FiberAbs01}
\end{center}
\end{figure}

\begin{figure}
\begin{center}
\includegraphics[trim=.7cm 0.5cm  0.6cm 0.0cm,clip,width=\hsize]{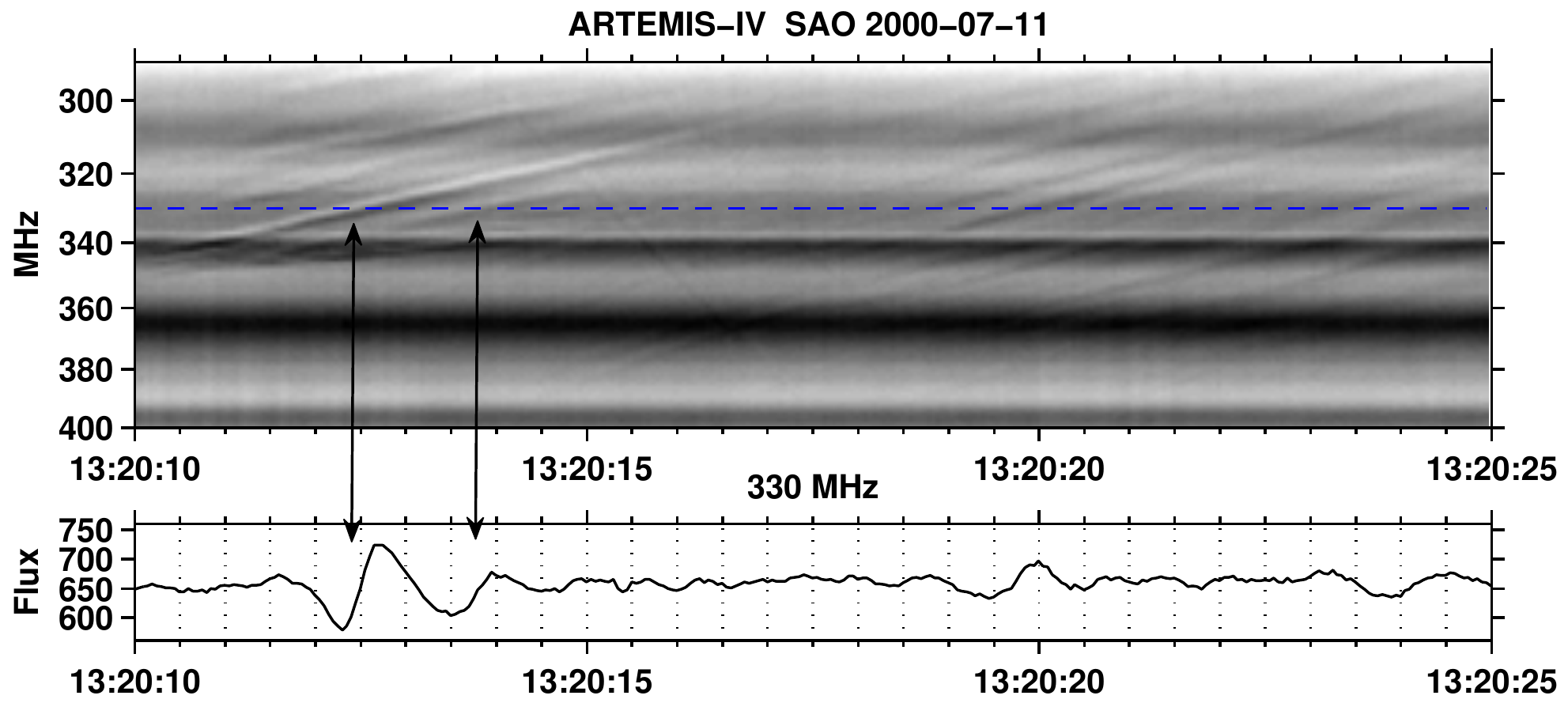}     
\includegraphics[trim=.7cm 0.0cm  0.6cm 0.0cm,clip,width=1.0\hsize]{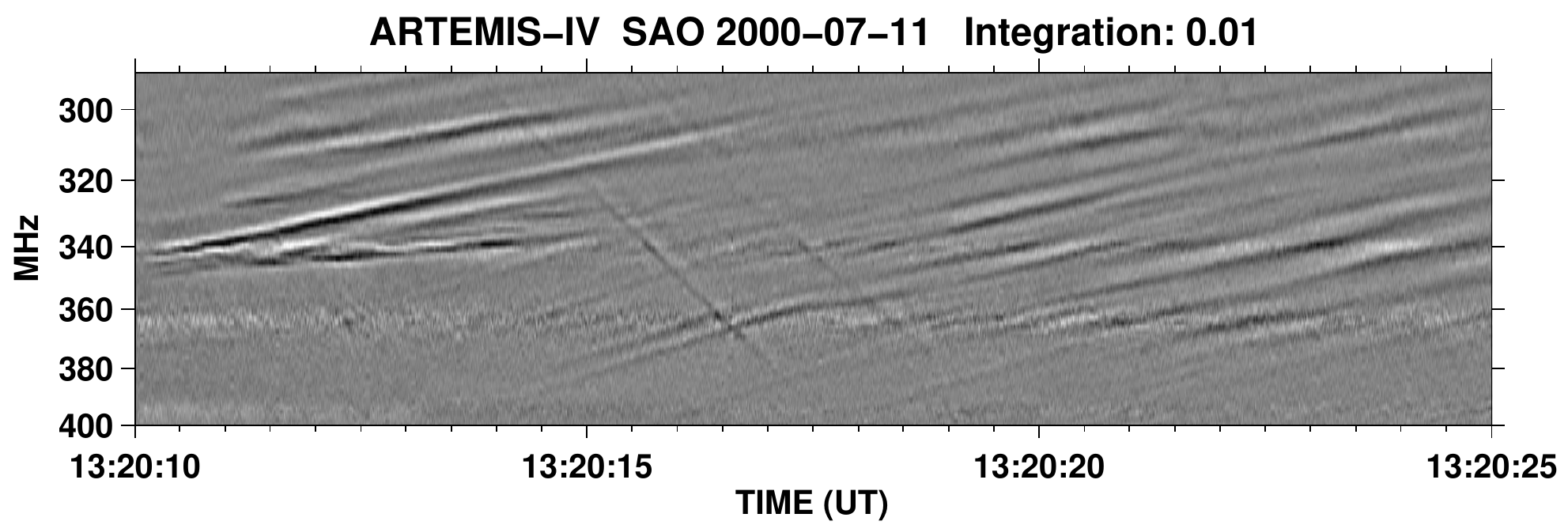}      
\caption{Typical fibers with an emission ridge and an absorption ridge at lower frequency. Top: Dynamic spectrum. Middle: Time profile at the frequency indicated by a dashed line on the dynamic spectrum, the arrows point to the absorption ridges of two consecutive fibers. Bottom: Filtered dynamic spectrum.}
\label{FiberNormal}
\end{center}
\end{figure}
\begin{figure}
\begin{center}
\includegraphics[width=\hsize, trim=0cm 0.0cm  0cm 0.2cm,clip]{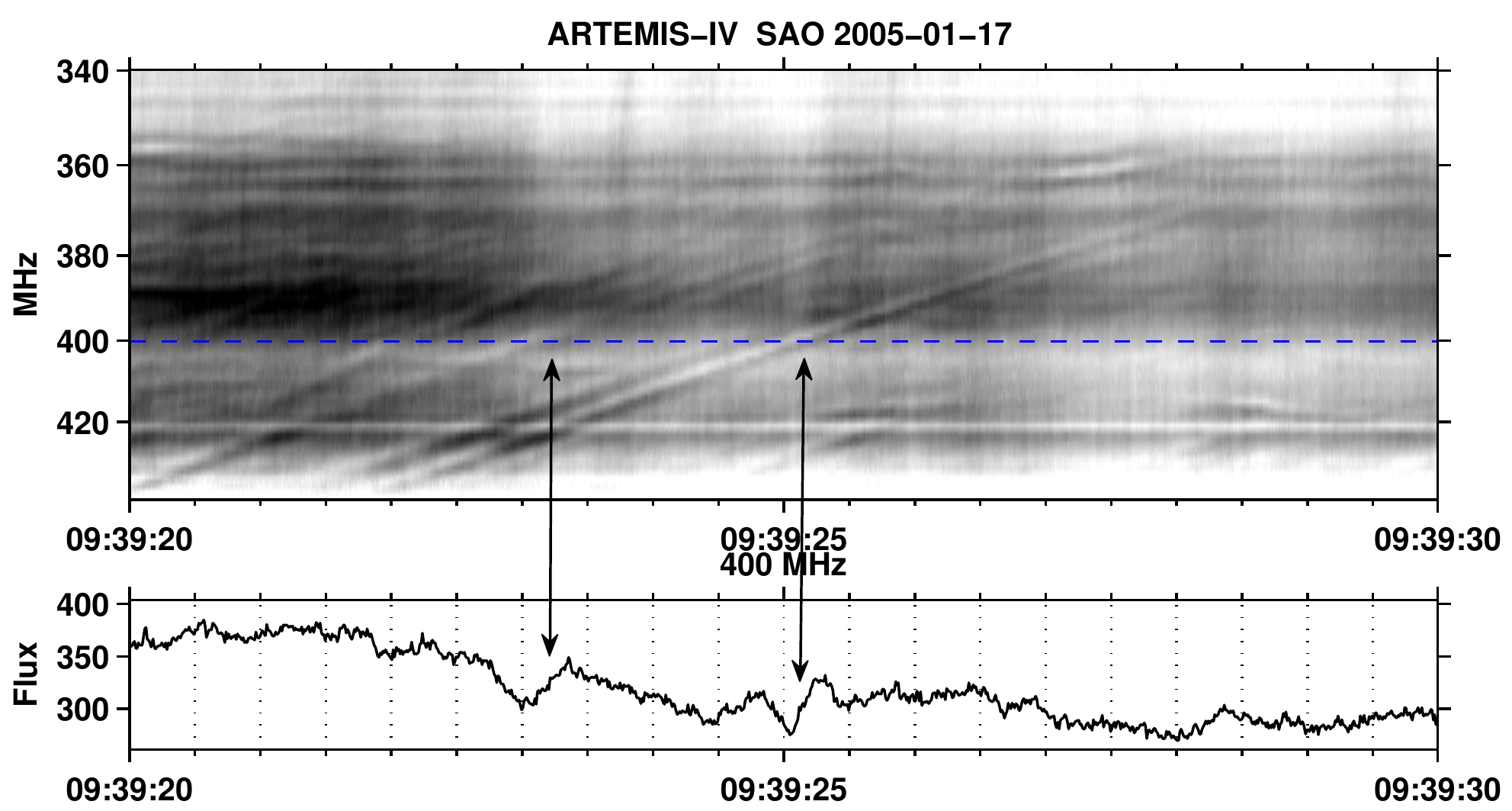}\\
\includegraphics[width=\hsize, trim=0cm 0.0cm  0cm 0.2cm,clip]{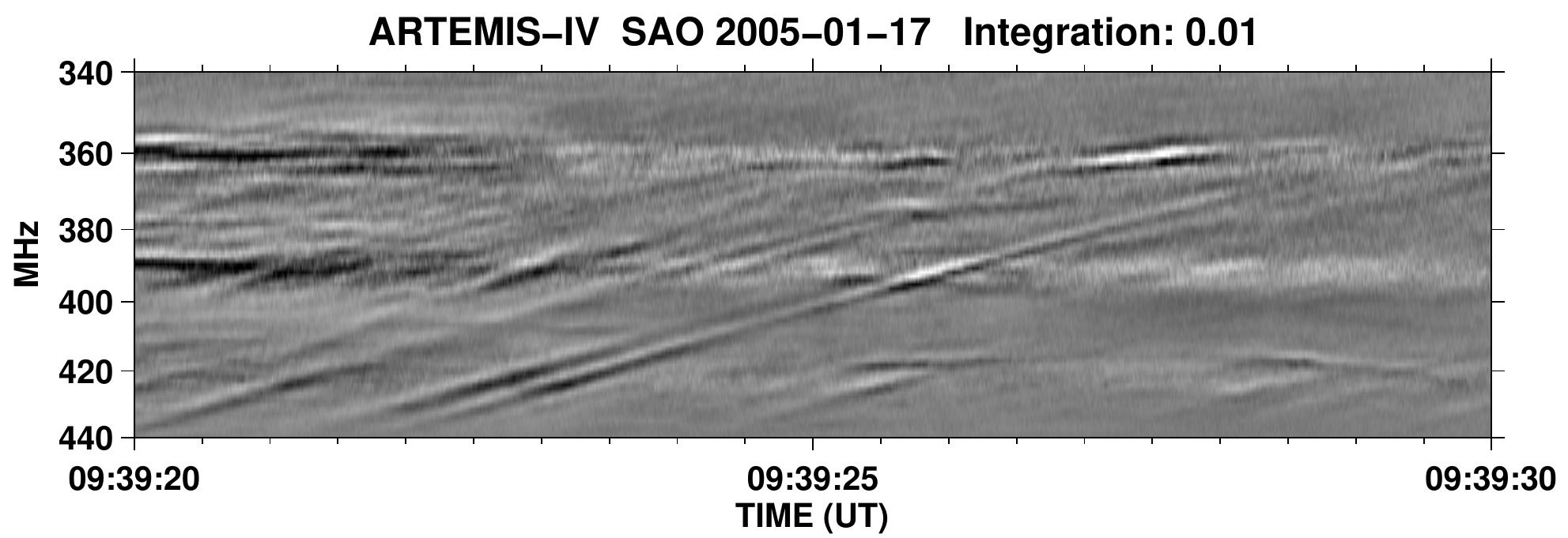}                                         
\caption{Upper panel: Fibers with absorption ridge and LF emission ridge in the interval of 09:39:20 to 09:39:30 UT; this is, in a sense, the opposite of figure \ref{FiberNormal}. Bottom: Time intensity profile at frequency indicated by dashed line; arrows point to two different fibers of the specific category.}
\label{FiberEmAbs01}
\end{center}
\end{figure}
\begin{figure}
\begin{center}
\includegraphics[width=\hsize]{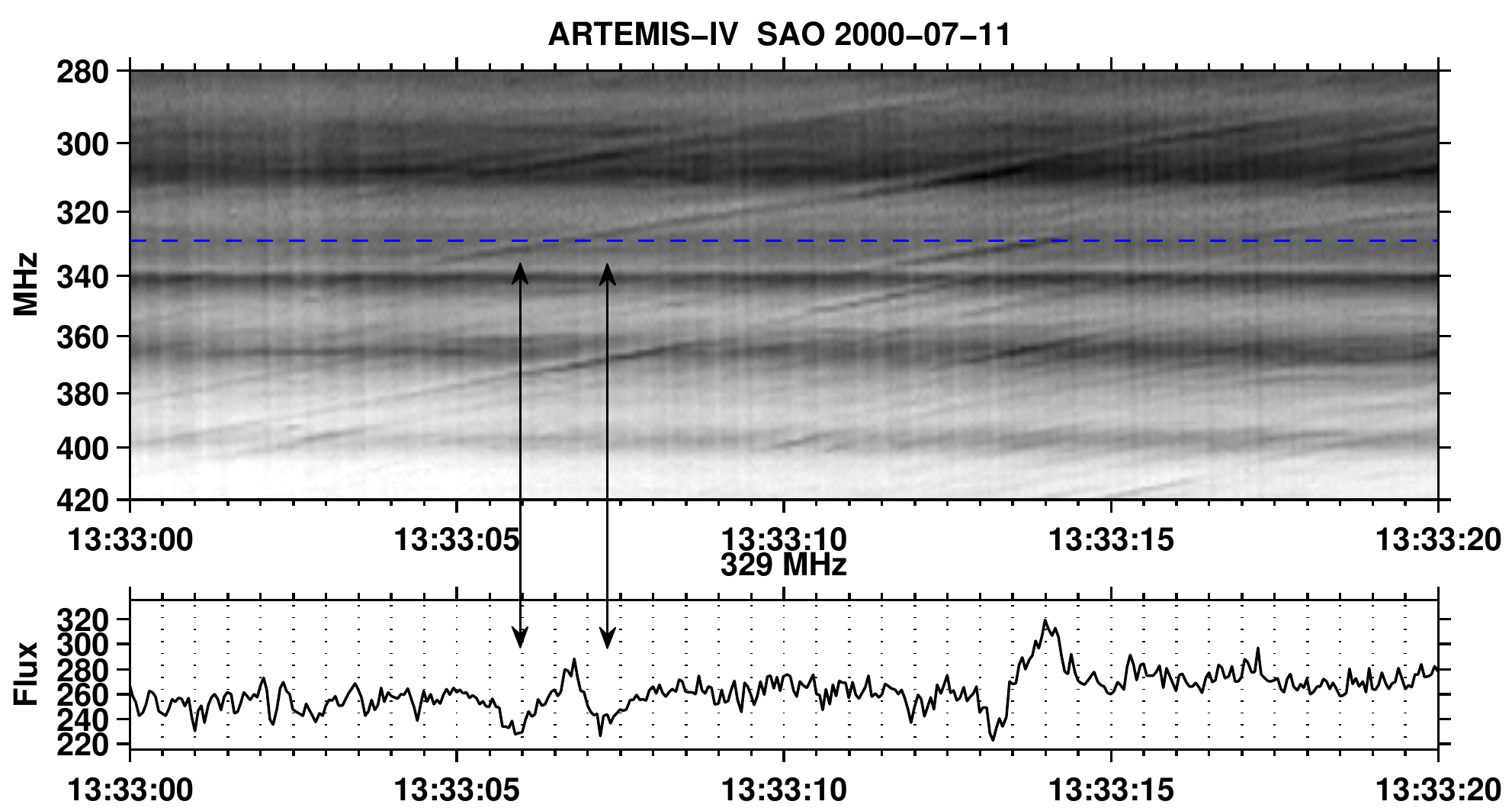}  
\includegraphics[width=\hsize,height=3cm]{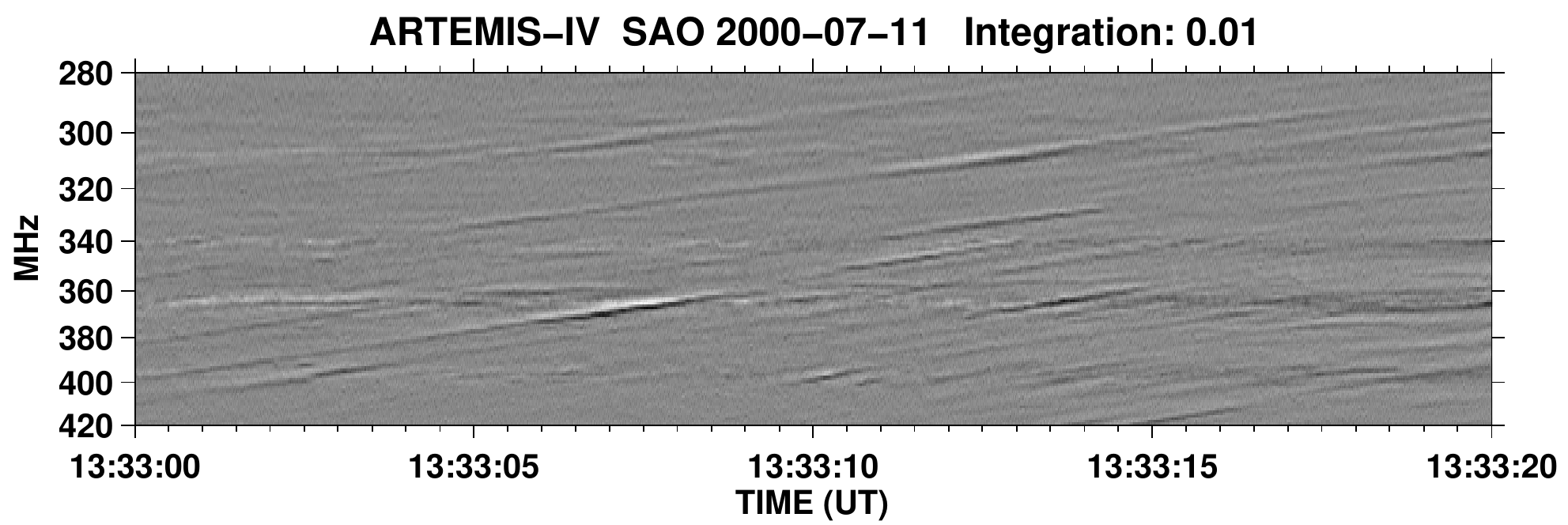}   
\caption{Fibers with an emission ridge between two absorption ridges.  Top: Raw dynamic spectrum. Middle: Intensity profile at the frequency marked by the dashed line; arrows point to the absorption ridges. Bottom: Filtered dynamic spectrum.}
\label{FiberAbsEmAbs}
\end{center}
\end{figure}
\begin{figure}
\begin{center}
\includegraphics[trim=1.2cm 0.4cm  1.2cm 0.0cm,clip,width=\hsize]{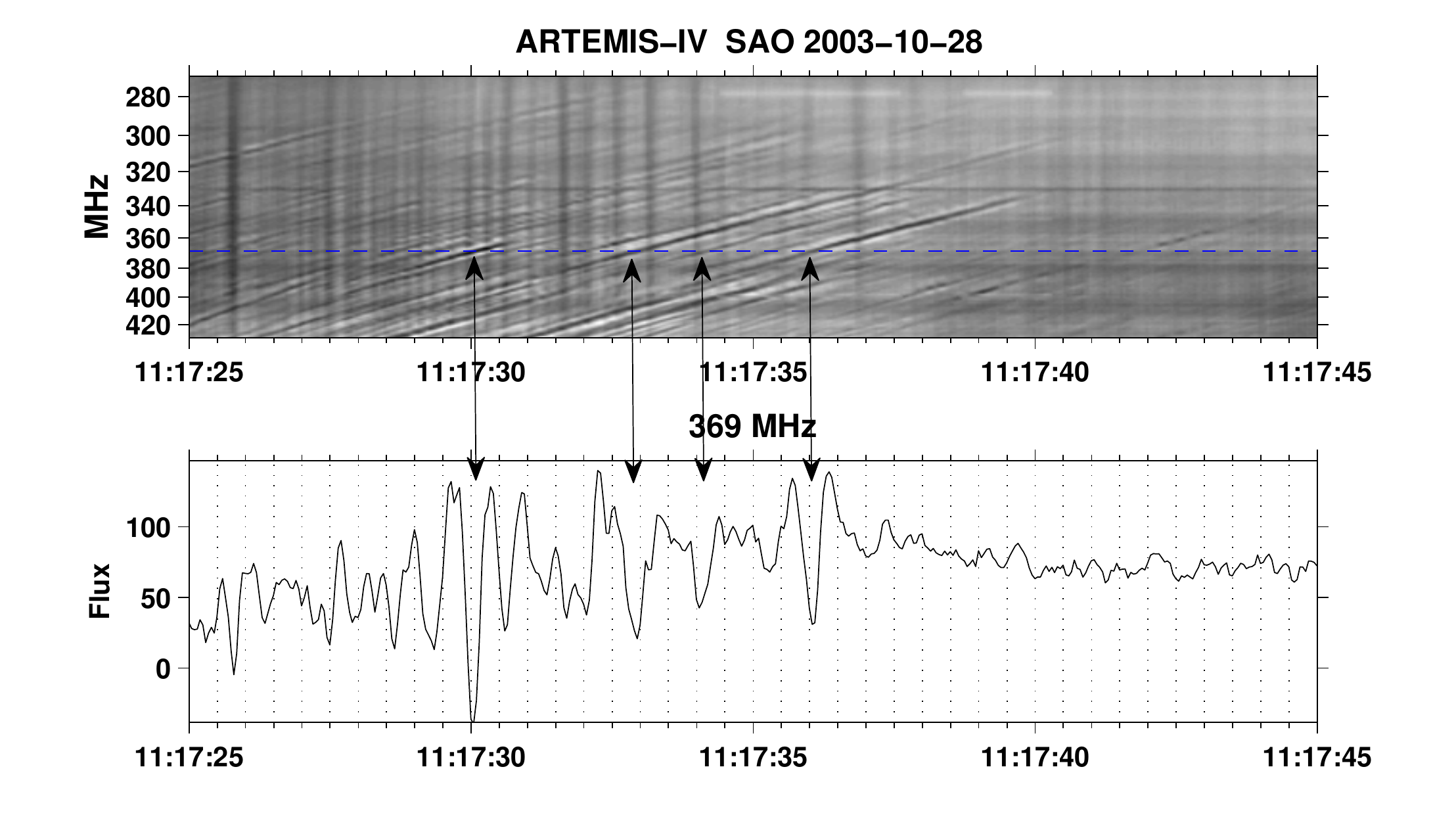}
\includegraphics[width=\hsize]{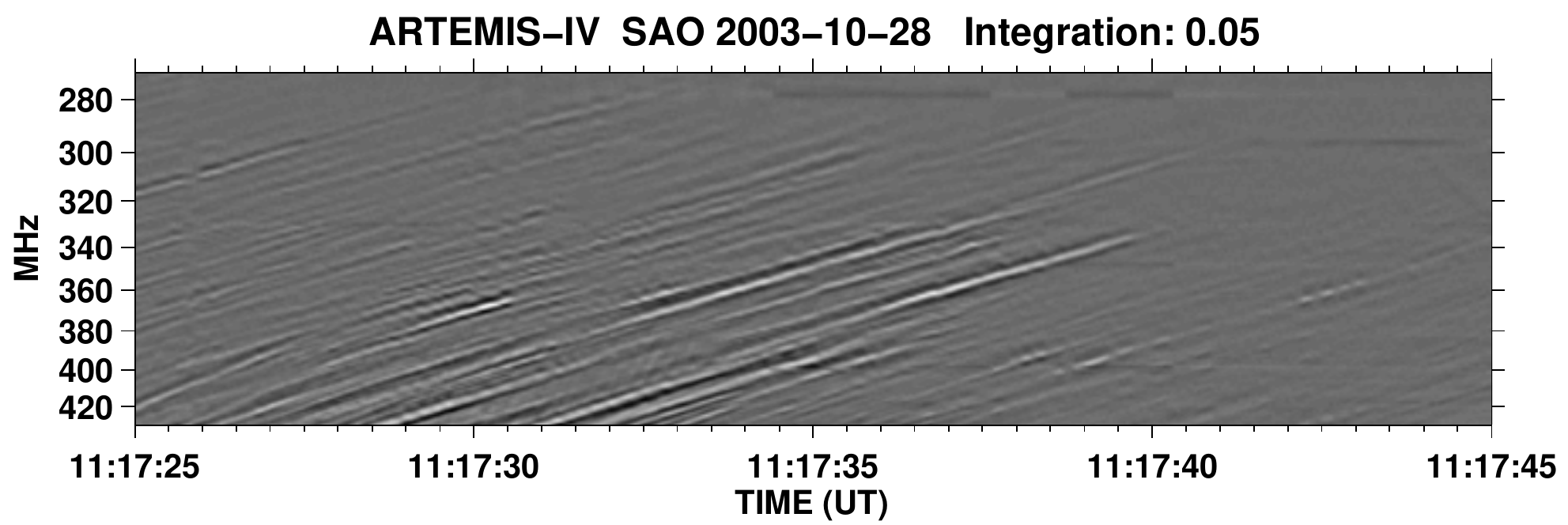}
\caption{Fibers with an absorption ridge between two emission ridges. Top: Dynamic spectrum (raw data). Middle: Time intensity profile at the frequency indicated by the dashed line; arrows point to the absorption ridges. Bottom: Filtered dynamic spectrum.}
\label{FiberEmAbsEm}
\end{center}
\end{figure}
\begin{figure}
\begin{center}
\includegraphics[trim=.7cm 0.0cm  0.6cm 0.0cm,clip,width=\hsize]{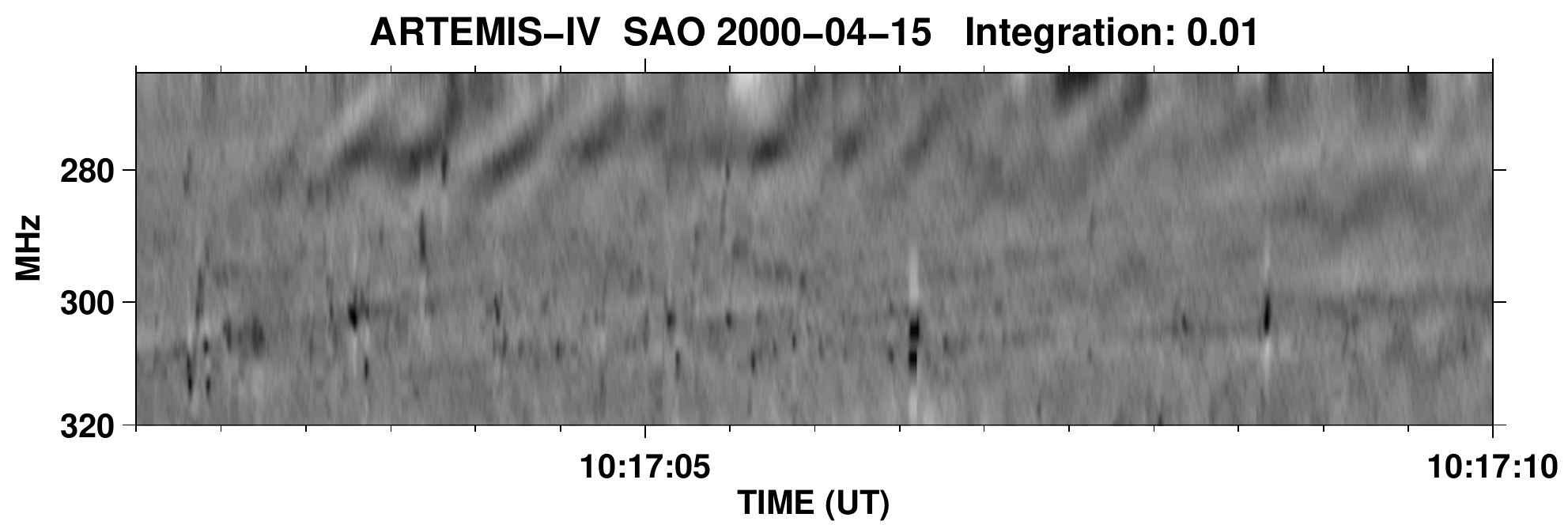}
\caption{Dynamic spectra of the single rope-like fiber chain ({\bf Event 40}) of our sample. This group of fast ropes, exhibit absorption ridge at lower frequencies, drift rate -33 MHz\,s $^{-1}$ (relative drift rate -0.12 s$^{-1}$),  instantaneous bandwidth 6 MHz (0.022 relative) and total frequency \mbox{extent  $\rm \simeq 20 MHz$.}}
\label{Fiberropes01}
\end{center}
\end{figure}
\begin{enumerate}
\item{Fibers with emission ridges only (Fig. \ref{fiberIsoEmission01}).}
\item{Fibers appearing in absorption on the Type IV continuum (Fig. \ref{FiberAbs01}).}
\item{Fibers with an emission ridge and a lower frequency (LF) absorption ridge (Fig. \ref{FiberNormal}).}
\item{Fibers with an emission ridge and high frequency (HF) absorption ridge (Fig. \ref{FiberEmAbs01}).}
\item{Fibers with an emission ridge between two absorption ridges (Fig. \ref{FiberAbsEmAbs}).}
\item{Two emission ridges separated by an absorption ridge \mbox{(Fig. \ref{FiberEmAbsEm}).}}
 \end{enumerate}
\begin{figure}
\begin{center}
\includegraphics[width=\hsize]{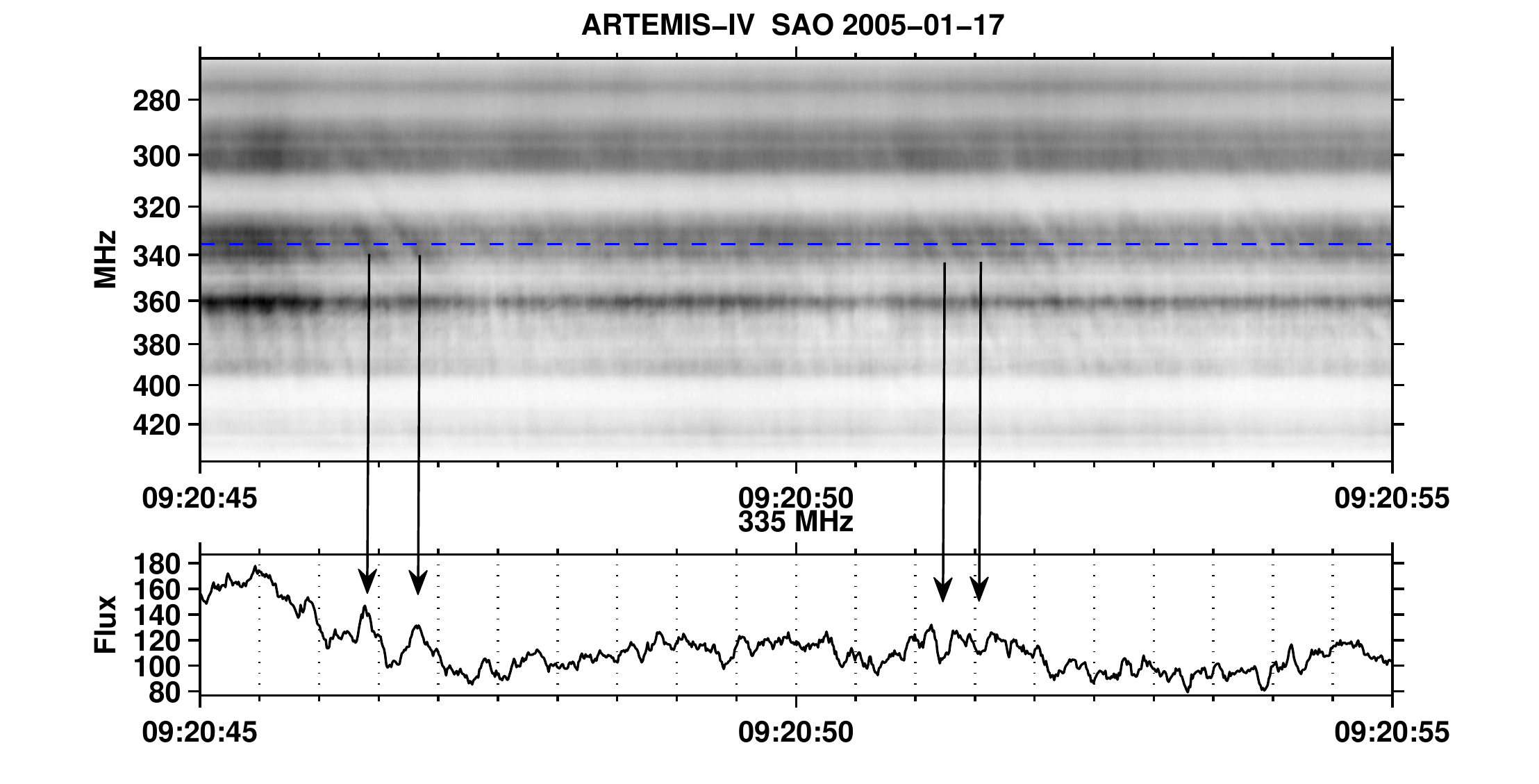}
\includegraphics[width=\hsize]{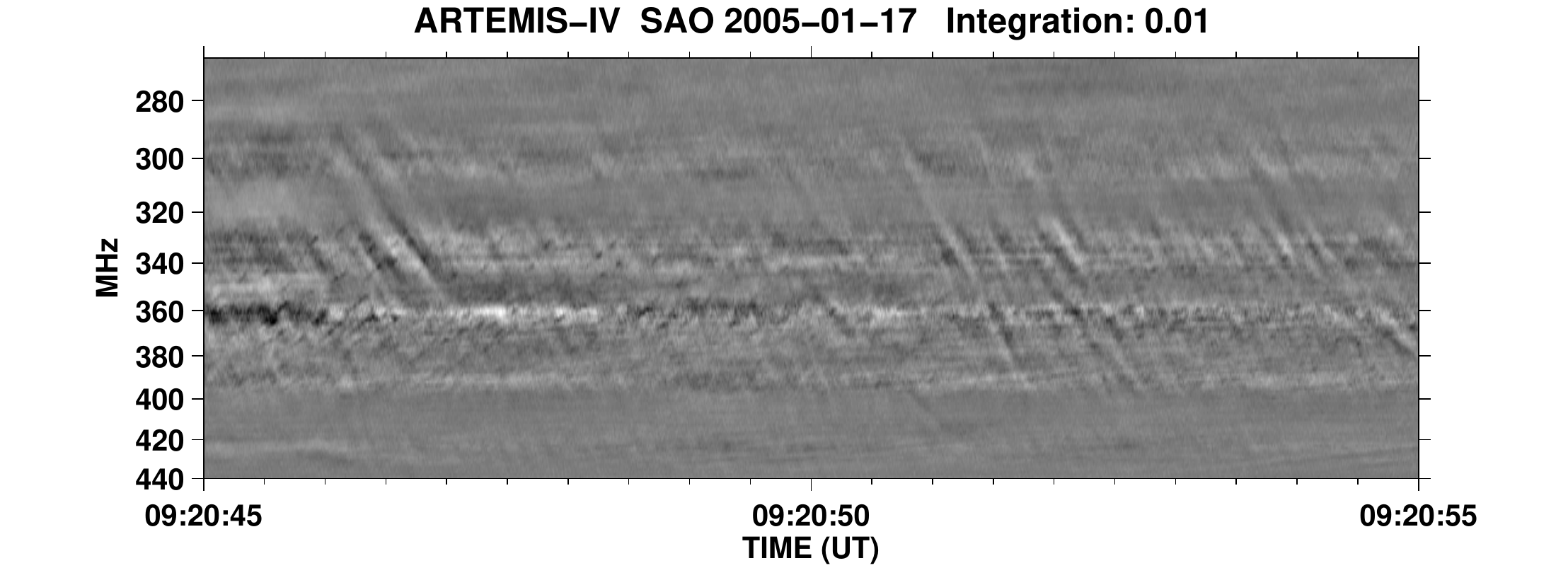}
\caption{Fast drift fiber bursts 17 January 2005 09:21 UT with a group of emission ridges and a group of absorption ridges on the Type IV continuum. Top:  Dynamic spectrum (raw, negative). Middle: Intensity profile at  335 MHz (dashed line in the top panel); arrows point to the  emission and absorption  ridges respectively. Bottom: filtered dynamic spectrum.}
\label{FDB44}
\end{center}
\end{figure}

Out of the six classes of typical fibers, the fibers with an emission ridge and a lower frequency (LF) absorption ridge are the vast majority in our data set. The remaining classes are represented by a few isolated cases.  

Interpretations based on the Langmuir wave--whistler interaction as regards the first four classes have have been published in the literature \citep[see reviews by~][~for example]{Chernov2006,Chernov2011} and are summarized in Sect. \ref{3W}. The 
{\bf remaining} two classes are discussed in this section with a tentative explanation.

The class of fibers that has an emission ridge between two absorption ridges was reported, probably for the first time,  by \citet{Messmer2002},  \citep[see also Figure 4.22 in][]{Chernov2011}. The class of fibers with two emission ridges separated by an absorption ridge, is first reported here to the best of our knowledge. This type of fibers appeared twice in our data set, on the 28 October 2003 fibers and the 14 July 2005 narrowband fibers.

These two groups probably result from combined coalescence ($\rm l+w\rightarrow t$) and decay ($\rm l\rightarrow t+w$). This indicates energy flow from the Langmuir wave background toward higher and lower frequencies creating an absorption ridge between two emission ridges. It may also lead to the opposite, where two absorption ridges are depleted transfering their energy to an emission ridge sandwiched between them. The question on the conditions that favor either mechanism is, to the knowledge of the authors, open.

\subsection{Rope-like fiber bursts} \label{Ropes}

The \RopeFibers~case  in our dataset (Event 40 in Table \ref{Dataset}) is presented in Fig. \ref{Fiberropes01} The individual ropes exhibit absorption ridge at lower frequencies, drift rate \mbox{ -33 MHz\,s $^{-1}$} (relative drift rate -0.12 s$^{-1}$),  instantaneous bandwidth 6 MHz (0.022 relative) and total frequency \mbox{extent  $\rm \simeq 20 MHz$.} No chain drift was observed in this case.

\begin{figure}
\begin{center}
\includegraphics[width=\hsize]{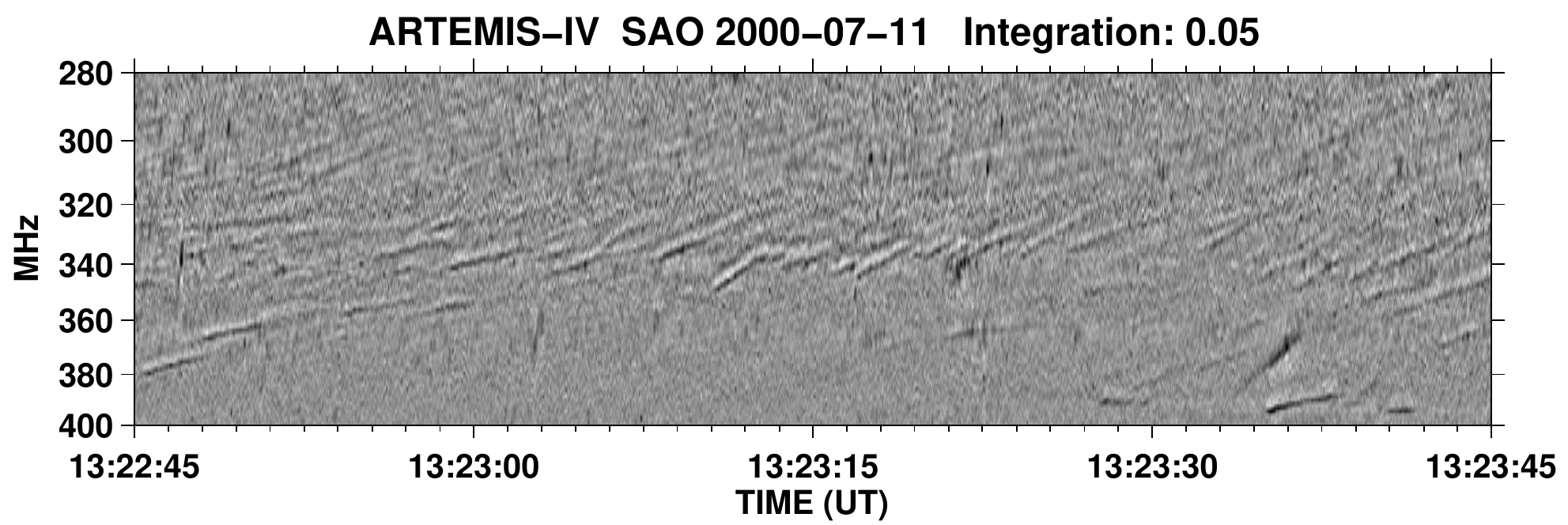}               
\includegraphics[width=\hsize]{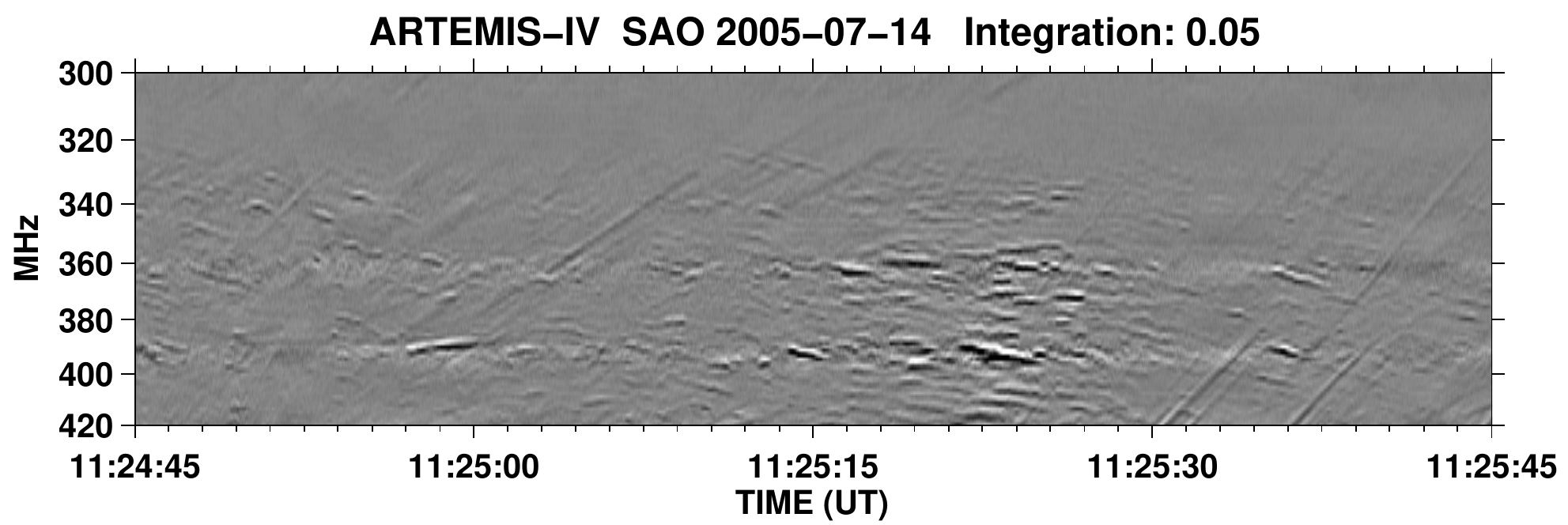}
\includegraphics[trim=.7cm 0.cm  0.6cm 0.0cm,clip=0,width=0.95\hsize]{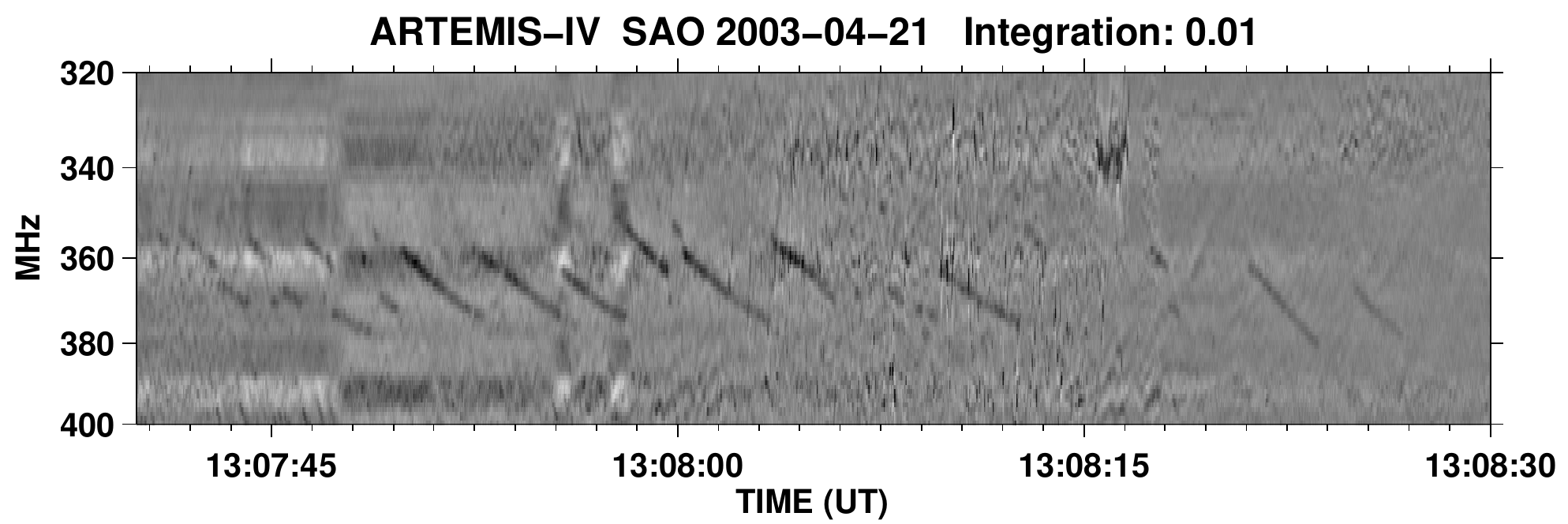}
\caption{Groups of narrowband IDBs (Events 45, 46, 47). Upper panel: 11 June 2000, narrowband fibers with drift rate in the range $-1.56$\,MHz\,s$^{-1}$ to $-6.56$\,MHz\,s$^{-1}$ and total bandwidth, $\Delta f_{\rm tot}\simeq$ 4-19\,MHz. Middle  panel: 14 June 2005, typical  fiber bursts overlapping with a cloud of mostly reverse-drift narrowband IDBs; their drift rate is $\simeq$0.45-2.46\,MHz\,s$^{-1}$, their  total bandwidth ($\Delta f_{\rm tot}\simeq 1$-6\,MHz). Lower Panel: 21 April 2003, reverse drift narrowband fibers with emission component only. Their drift rate is $\rm 4.8 MHz\,s^{-1}$ and the  total \mbox{bandwidth $\rm \simeq 5 MHz$}}
\label{Fibernarrow}
\end{center}
\end{figure}

\subsection{Fast drift fiber bursts} \label{FDB}

The fast drift bursts, which are provisionally included in \citet{Bouratzis2015} as {\bf an intermediate drift burst subcategory, exhibit characteristics consistent with the classification scheme of this section.} Despite their drift rate, which is comparable to the type III bursts, they do have emission and absorption ridges as presented in Fig. \ref{FDB44}. 

\subsection{Narrowband intermediate drift bursts} \label{NarrowFiber}
Bursts with narrow frequency extent  ($\approx$ 10 MHz on average, or  3\% relative) were recorded in our dynamic spectra in \NarrowFibers~cases (Events 45, 46, and 47 in Table \ref{Dataset}). In the classification of fine structure \citep[as described by][]{Bouratzis2015}, we consider such events to be members of intermediate drift bursts class, instead of members of the narrowband burst family for two reasons. Firstly, they exhibit at least an absorption ridge  at the lower-frequency side  (Events 45, 46) with a number of narrowband fibers in the Event 47 group exhibiting emission-absorption-emission structure; in all three cases this is typical characteristic of fiber bursts. Secondly the frequency drift rate of narrowband events is comparable (even larger) with the type III bursts drift rate \citep{Bouratzis2016}, which is not the case of  the narrowband fiber bursts whose drift rate is almost equal or smaller  than the typical intermediate drift bursts.

The narrowband IDBs of our data set are presented in the \NarrowFibers~panels of Fig. \ref{Fibernarrow} and exhibit a significant diversity. In the top panel of the figure the narrowband IDBs, recorded on the 11 July 2000, have frequency drift rate in the range of  $\rm  -1.56~MHz\,$s$^{-1}$ to $\rm  -6.56~MHz\,$s$^{-1}$ with total frequency extent ($\Delta f_{\rm tot}$) from 4 MHz to 19 MHz. In the middle  panel  there are typical  fiber bursts overlapping with with  a cloud of narrowband IDBs with, mostly, positive drift rate of 0.45-2.46~MHz$\,$s$^{-1}$ and  total bandwidth ($\Delta f_{\rm tot}$) from 1 MHz to 6 MHz; these were obtained on the 14 July 2005. The bottom panel depicts the last group, observed on 21 April 2003, which consists of reverse drift narrowband fibers with emission component only. Their drift rate is $\rm 4.8 MHz\,s^{-1}$ (relative drift rate $\rm 0.013s^{-1}$) and the total frequency extent is $\rm \simeq 5 MHz$. Some fibers of the  last group exhibit some similarity to  the broken quasi-fibers (due to absence of absorption ridge) reported by \citet{Kuijpers&Slottje1976}.

\subsection{Complex fiber groups} \label{ChainBurst}
\begin{figure}[!h]
\begin{center}
\includegraphics[width=\hsize]{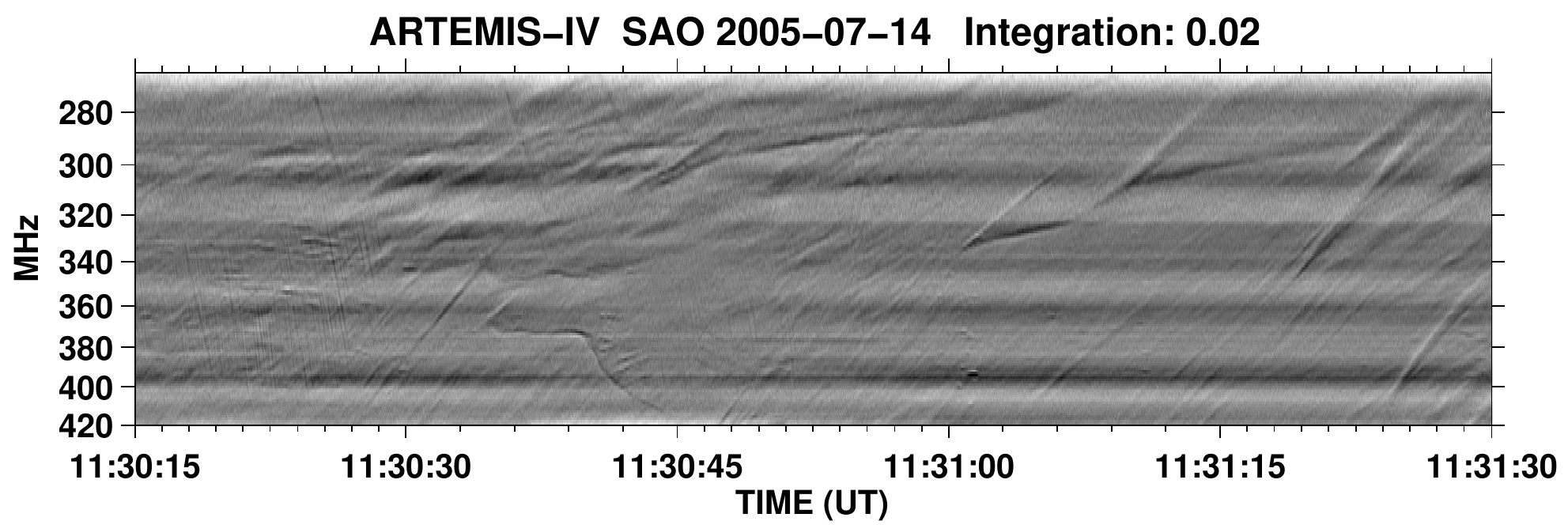}
\caption{Dynamic spectrum of a complex group of fibers.}
\label{Fiberpecul02}
\end{center}
\end{figure}

Some complex groups of fibers with different drift rates, are found to overlap on dynamic spectra; is unclear whether this configuration might constitute a separate subclass; see Fig. \ref{Fiberpecul02}, for an example.
The fiber repetition period, for this type of fiber, was in the range 0.46--2.3 s. The individual fibers had duration at fixed frequency $\rm \approx0.4 s$ and instantaneous bandwidth of \mbox{$\approx$~3.5 MHz}; the frequency extent (total bandwidth) was found, on average, \mbox{$\Delta f_{\rm tot}\approx $40MHz} which corresponds to \mbox{$\Delta f_{\rm tot}/f\approx$ 10\%}. The majority of the fibers have negative drift rates $ df/fdt\approx-{\rm 0.023~s^{-1}}$ while those with positive drift reach rates $ df/fdt\approx{\rm 0.034s^{-1}}$; the latter are always coexistent with fibers of negative drift rate. Some of the fiber structures in Fig.  \ref{Fiberpecul02} are accompanied by an unusually slowly drifting branch so they may qualify as slow drifting fibers \citep[see Sect. 4.4.4.1 of ][]{Chernov2011} despite their low total duration on the order of tens of seconds.

\begin{figure}
\begin{center}
\begin{tabular}{cc}     
\includegraphics[width=0.23\textwidth]{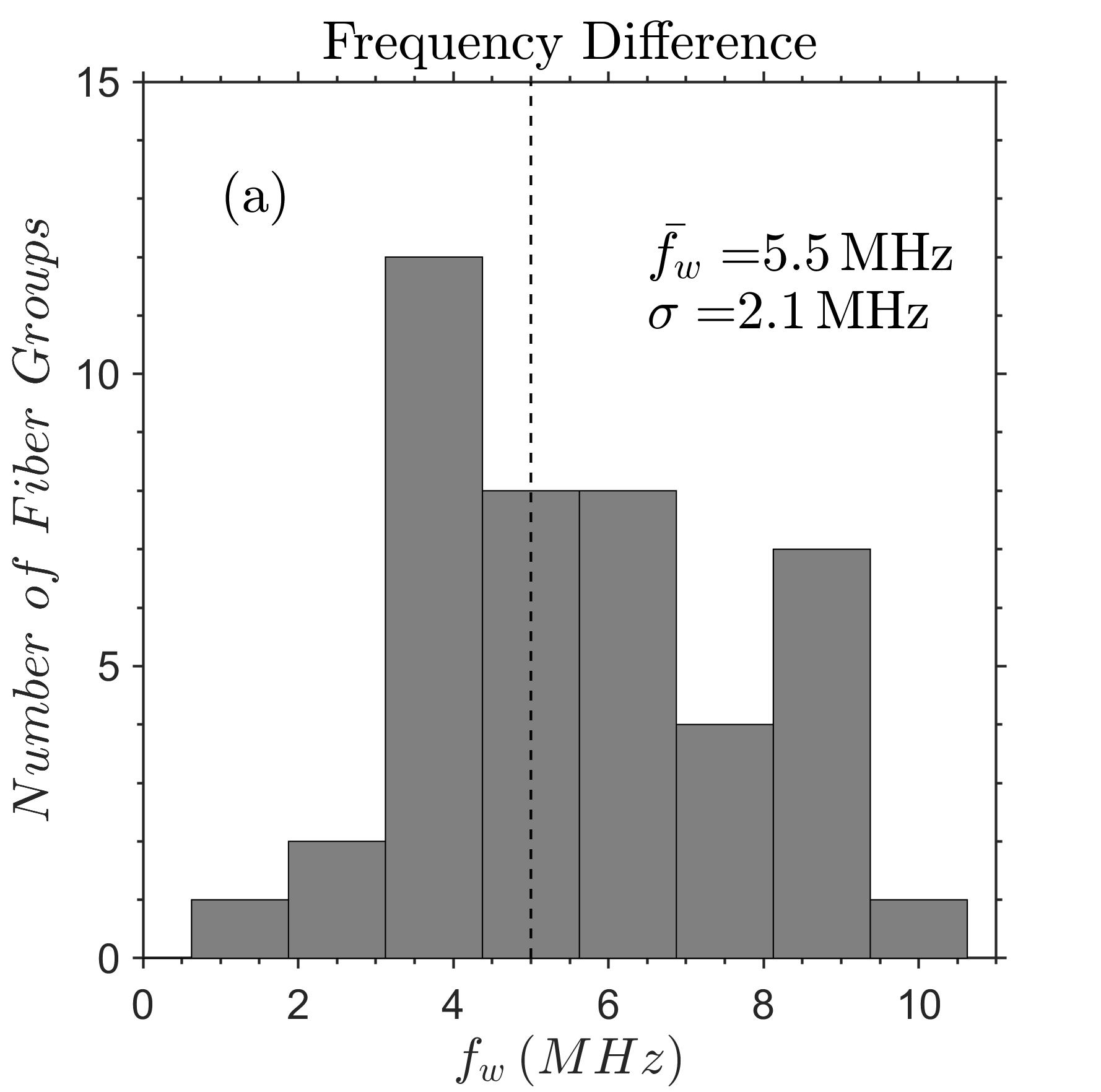}             & \includegraphics[width=0.23\textwidth]{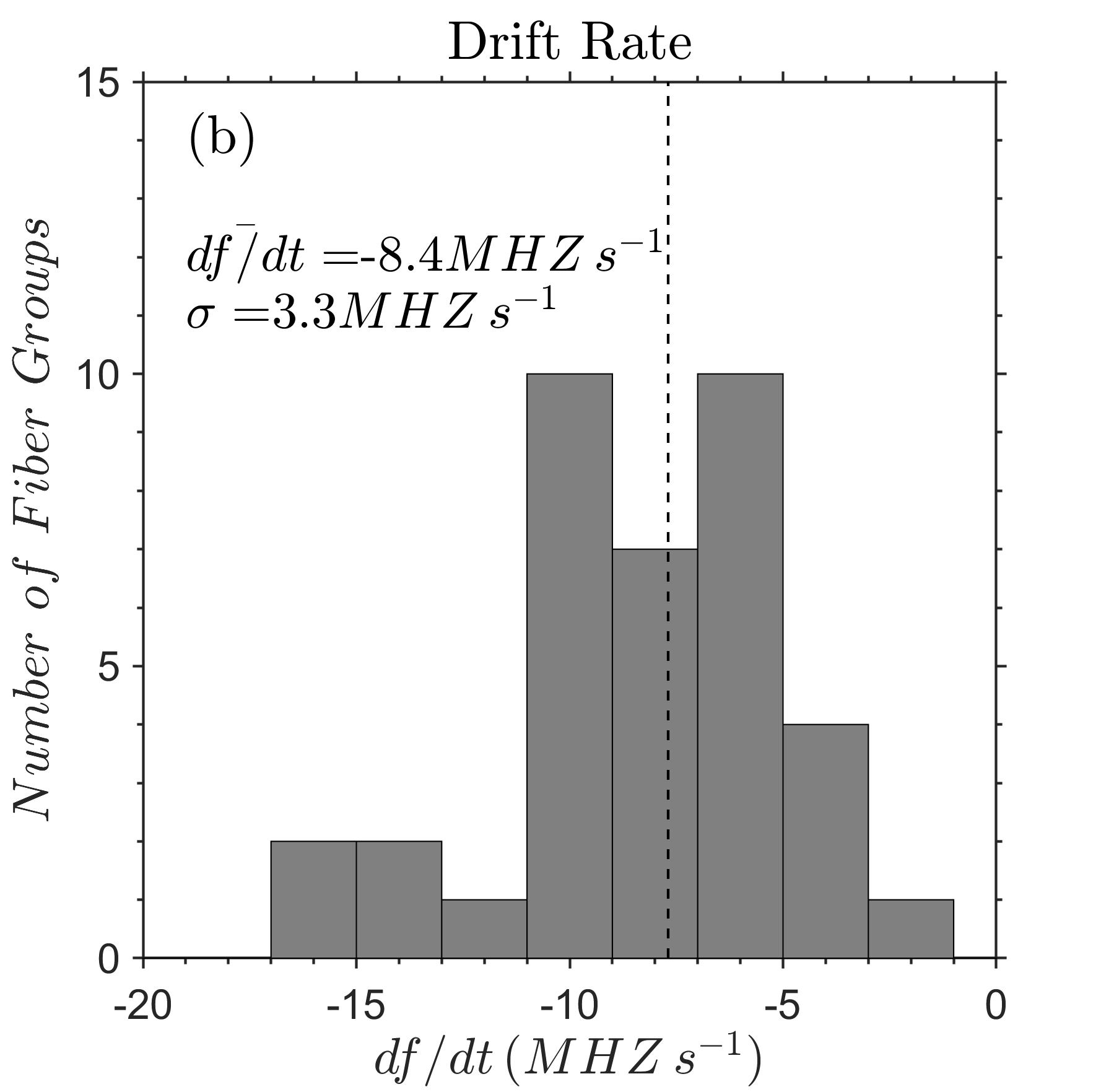} \\        
\includegraphics[width=0.23\textwidth]{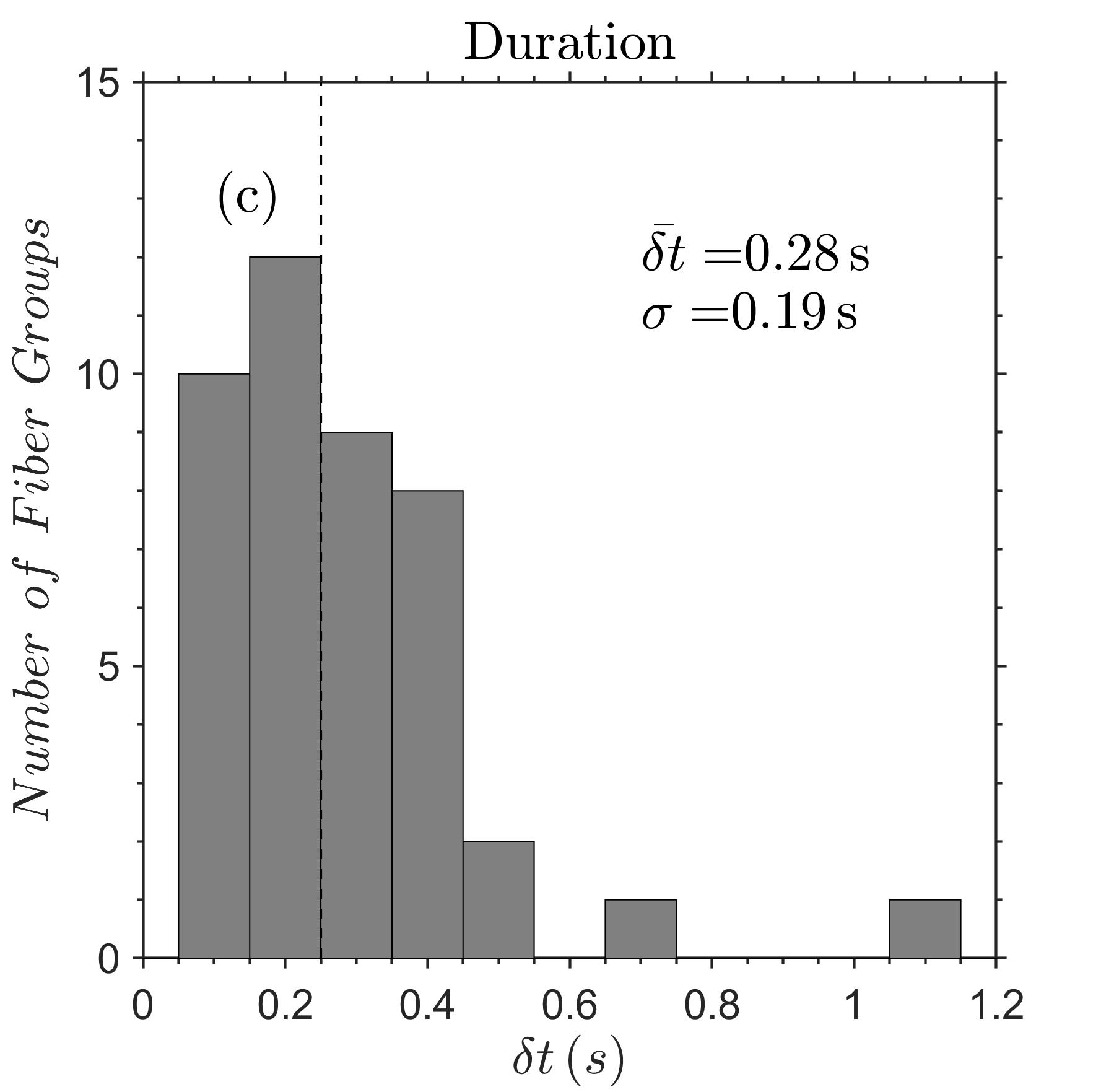}       & \includegraphics[width=0.23\textwidth]{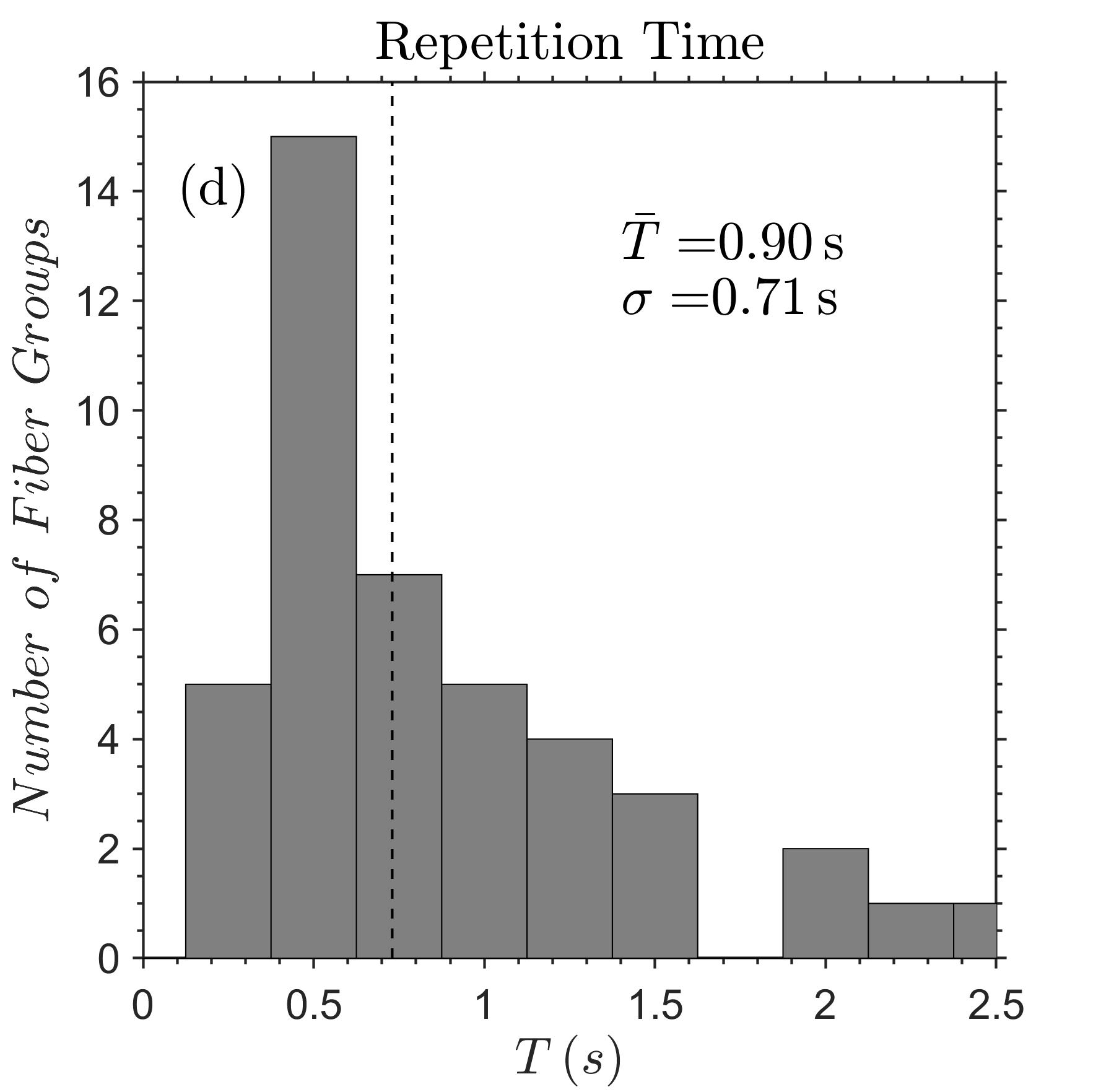}
\end{tabular}
\caption{Distribution of fiber parameters computed by 2D autocorrelation. (a): Frequency difference between absorption-emission peaks (see Fig. \ref{2Dauto}d), (b): Frequency drift rate, calculated for the outbound typical fibers (groups 1-37). (c): Duration at fixed frequency. (d): Average repetition time. Both (c) \& (d)  were calculated at the central frequency for each group.  In all panels except (b) data from groups 1-44 of Table \ref{Dataset} were used, excluding narrowband fibers.}
\label{distBW_Dur_Per}
\end{center}
\end{figure}
 \begin{table*}
\centering
\caption{Average Properties of Individual Intermediate Drift Bursts (Direct Measurement) and Groups analysed by means of 2D Auto-Correlation.}
\label{TableFibers}
\begin{tabular}{p{2.2cm} ll ll p{1.8cm} ll }
\hline
																	& \multicolumn{2}{c}{Typical Fibers}		&  \multicolumn{2}{c}{Fast Drift Bursts} 			& Rope-like fibers	& \multicolumn{2}{c}{{Narrow-band IDBs}}\\
\hline
Fiber Groups														& \TypicalFibersNeg	& \TypicalFibersPos		&\FastDriftFibersNeg	&	\FastDriftFibersPos		&	\RopeFibersNeg	  &	{\NarrowFibersNeg}	&{\NarrowFibersPos}\\
\hline
Freq. Drift															& Negative  		& Positive  			& 	Negative  			& 	Positive  				& Negative  		  & Negative  		& Positive 				\\
{\Large$\frac{df}{dt}$} \mbox{\rm \small(MHz$\rm \cdot s^{-1})$} 	& -8.43  $\pm$ 3.29	& 8.15 					&	-142.8 				&	73.4 $\pm$ 31.0			&	-45.3 			  &	-5.86 			&	3.71 $\pm$ 2.6		\\[0.2cm]
{\Large$\frac{df}{fdt}$} \mbox{\small (s$^{-1}$)}   			& -$\rm (20\pm8)10^{-3}$&$20\cdot10^{-3}$	&	-0.36 				& 	$\rm(191\pm84)10^{-3}$ 	&	-0.10 			  & -0.014			&	0.009 $\pm$ 0.006	\\[0.2cm]
$\delta t $ (s)            			 								& 0.30 $\pm$ 0.20 	& $24\cdot10^{-2}$		& $6\cdot10^{-2}$		&	$(9\pm2)\cdot10^{-2}$ 	& 	0.21  			  & 0.30			&	0.58 $\pm$ 0.39		\\
$T$ (s)						 										& 0.98 $\pm$ 0.74	& 0.60 					&	0.24				&	$(18\pm1)\cdot10^{-2}$ 	& 	0.43 			  & 0.80			&	0.80				\\

$\delta f$ {\small (MHz)}       							& 2.4 $\pm$ 0.8		& 2.4					&	4.0					& 	4.3 $\pm$ 1.2			& 	6.0			  	  & 2.0			    &	3.0 $\pm$ 1.4		\\
{\Large$\frac{\delta f}{f}$} ({\small$10^{-3}$})			& 9.0 $\pm$ 4		& 8.2 					&	13.2 				& 	12.7 $\pm$ 2.2			&	25.1 			  & 9.4			&	9.1 $\pm$ 4.8			\\[0.2cm]
$f_{\rm w}$ {\small (MHz)}       							& 5.5 $\pm$ 2.2		& 6.4					&	7.5					& 	5.9 $\pm$ 1.4			& 	6.90			  & 3.18			&	3.33 $\pm$ 1.65		\\
{\Large$\frac{f_{\rm w}}{f}$} ({\small$10^{-3}$})			& 15 $\pm$ 6		& 18 					&	25 					& 	17 $\pm$ 5				&	20 			  	  & 9				&	9 $\pm$ 5			\\[0.2cm]

\hline
Individual Fibers             			  							& 441				& 13					&	11					&	61						&	13				  &	19				& 31					\\
\hline
$\Delta t_{\rm tot}$ (s)				  							& 4.26 $\pm$ 2.61 	& 3.73 $\pm$ 2.29		& 	0.55 $\pm$ 0.17		&	1.04 $\pm$ 0.47 		& 	0.49 $\pm$ 0.11	  & 2.66 $\pm$ 1.10	& 2.0 $\pm$ 0.80		\\
$\Delta f_{\rm tot}$ {\small (MHz)}									& -37.9 $\pm$ 16.4	& 33.8 $\pm$ 21.3		&	-44.3 $\pm$ 19.1	& 	67.4 $\pm$ 14.0			& 	16.8 $\pm$ 2.9	  & 7.9 $\pm$ 4.3	& 8.4 $\pm$ 6.0			\\
{\Large$\frac{\Delta f_{\rm tot}}{f}$} ({\small$10^{-3}$})		  	& 103 $\pm$ 44		& 92 $\pm$ 58	 		&	138 $\pm$ 66		& 	198 $\pm$ 42			&   60 $\pm$ 10	  	  & 23 $\pm$ 13	 	& 23 $\pm$ 16			\\[0.2cm]
\hline
\end{tabular}
\end{table*}
\begin{figure}
\begin{center}
\begin{tabular}{cc}     
\includegraphics[width=0.45\hsize]{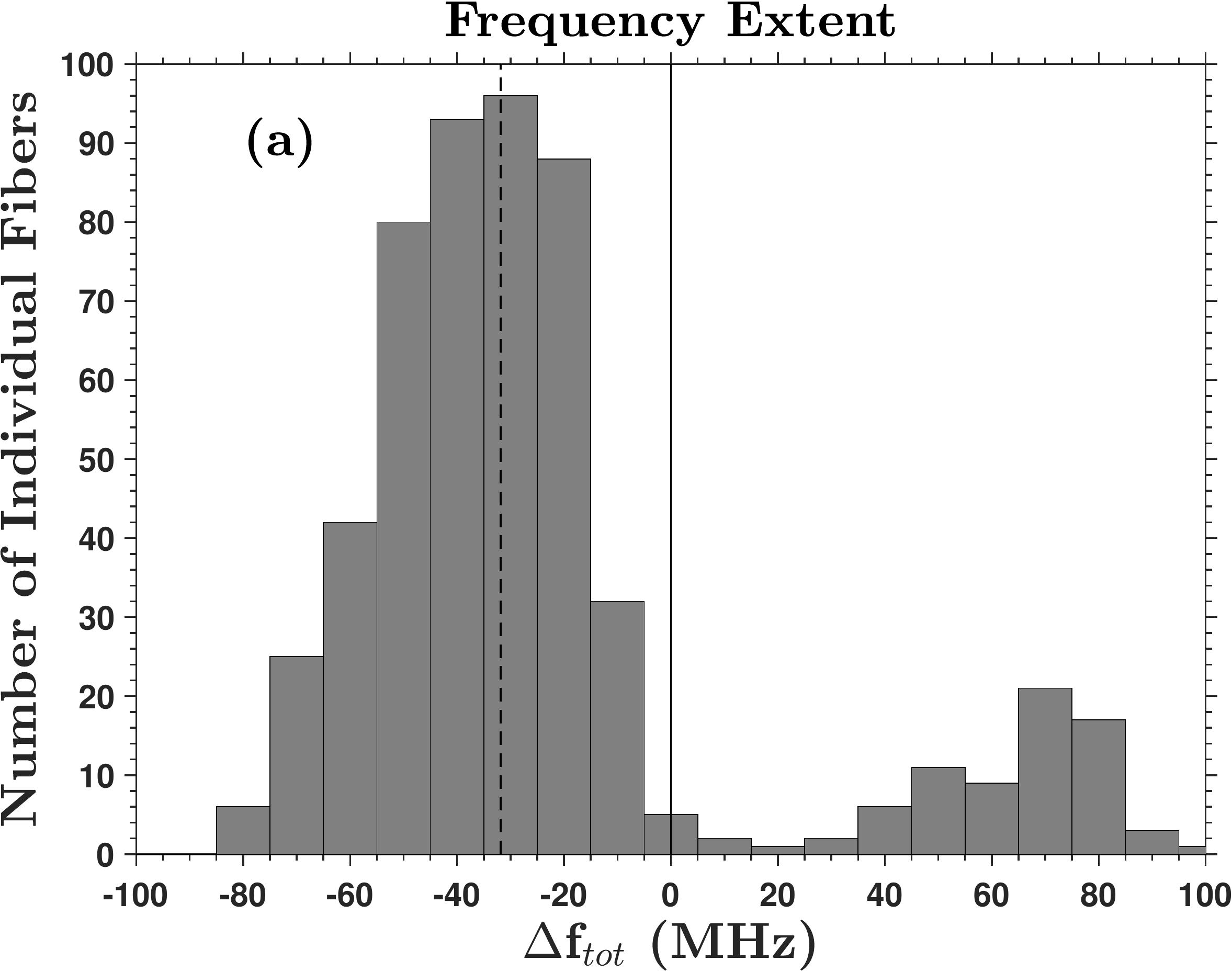}  & \includegraphics[width=0.45\hsize]{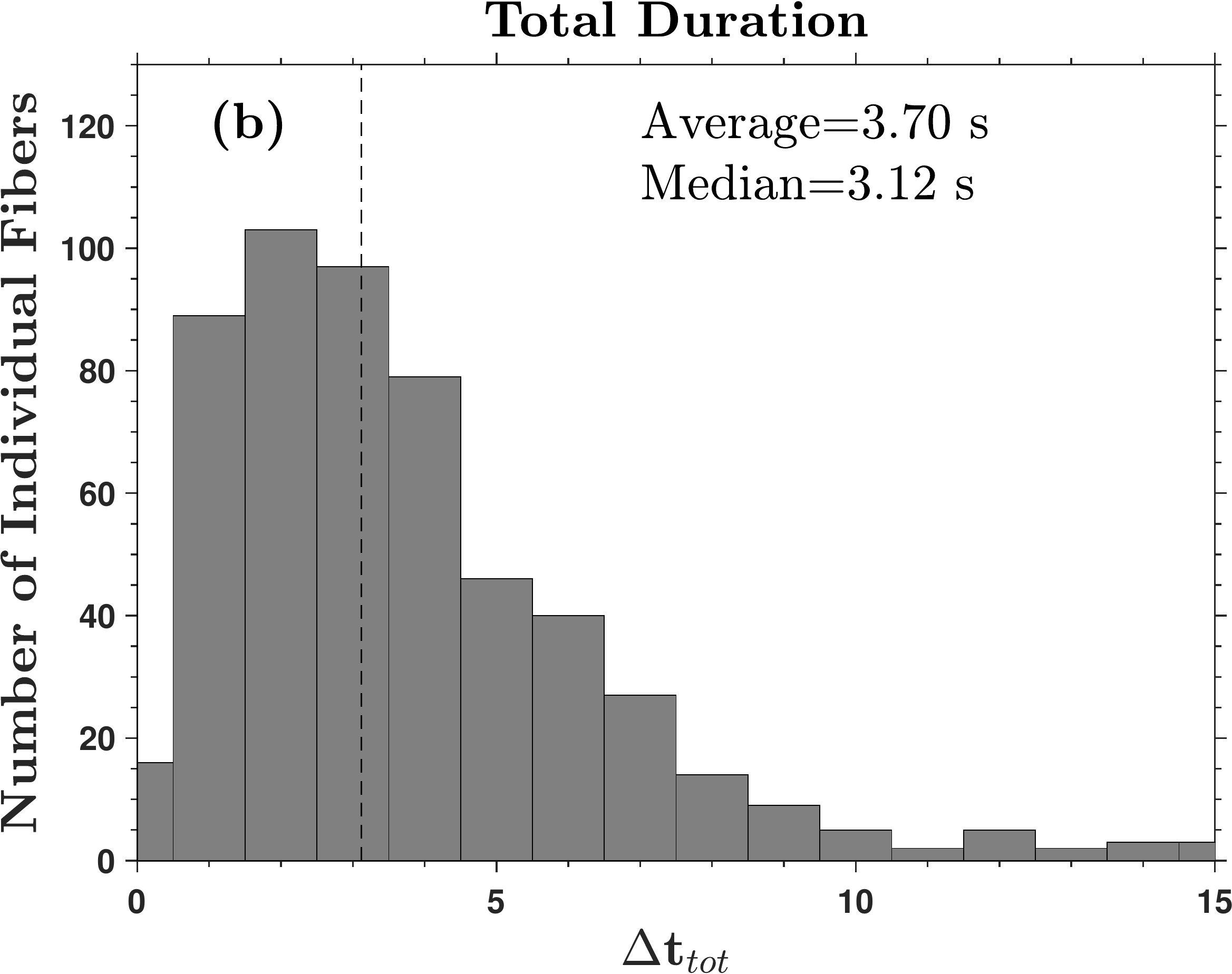} \\  
\end{tabular}
\caption{Distribution of characteristics of individual fibers (540 samples), measured directly from the dynamic spectra. (a) Frequency range, $\Delta f$. The negative values of bandwidth are from outbound bursts. (b): Total duration, $\Delta t_{\rm tot}$. The histogram bin near zero
in both panels
corresponds to narrowband fibers, such as depicted in Fig. \ref{Fibernarrow}.}
\label{distBW_Dur_SingleFiber}
\end{center}
\end{figure}
\section{Statistics}\label{SingleBurst}
The measured bulk parameters of fiber groups (frequency difference between absorption-emission peaks, frequency drift rate, duration at fixed frequency and repetition rate), obtained from 2D autocorrelation are given in Table \ref{Dataset} and the corresponding histograms are plotted in Fig. \ref{distBW_Dur_Per}. The measured characteristics of the 540 individual fiber bursts (total duration and frequency extent) are presented in Fig. \ref{distBW_Dur_SingleFiber}. The average values (with standard deviation) both for groups and individual fibers are presented  in Table \ref{TableFibers}. In addition to the drift rate in \mbox{MHz\,s$^{-1}$}, we include  the logarithmic drift rate in s$^{-1}$ within this table in order to facilitate comparison with results in different frequency ranges. We note that the fibers with two absorption or emission ridges which correspond to classes 5 and 6 of the typical fibers in Sect. \ref{TypicalFibers} are represented by few isolated cases, hence our measurements of the bulk parameters were not affected.

In our sample, most of the fibers had negative drift rates (outbound), with only one inbound group of typical fibers, {\bf two of narrowband IDBs} and three of FDBs; inbound fiber groups have not been recorded without co-existing outbound. On the average our results are consistent with similar statistics in the decimetric range \citep[see~][~their Table 2]{Fernandes2003}. Our fiber bursts exhibit logarithmic drift rates with mean value \mbox{$df/fdt\approx-0.027$s$^{-1}$}, quite similar to previous results \citep{Young61,Elgaroy73,Benz98}

The drift rate of FDBs, on the other hand,  \mbox{reach $\approx-0.35$s$^{-1}$}, well within the reported range of the type III bursts drift rate \citep[see ][]{Young61,Elgaroy73,Jiricka01,Jiricka02,Meszarosova05,nishimura2013}; the FDBs represent outliers within our sample of the intermediate drift bursts in Table \ref{TableFibers}.  Fiber bursts in emission and in absorption with high drift rate ($df/fdt\approx{\rm 0.15s^{-1}}$) in the decametric range \mbox{(10--30 MHz)} have been reported by \citet{Melnik2010}. The rope-like  intermediate drift burst family in Sect. \ref{Ropes}, \citep[see also ][]{Mann1989, Chernov2008, Chernov2011} is another borderline case.

In Fig. \ref{distBW_Dur_Per}(a) we present the distribution of the frequency difference between the absorption and emission peaks  ($f_w$); this is in the range 3.0-9.5\,MHz with  average of about 5.6\,MHz. The distribution of the    frequency difference normalized to the observation frequency ($f_w/f$) is depicted in Fig. \ref{distBW_Dur_Per}(b).

We measured the instantaneous bandwidth ($\delta f$) of the emission ridge as the full width at half maximum (FWHM). The mean value was 2.4\,MHz while the relative instantaneous bandwidth ($\delta f /f$) was $9.0 \cdot 10^{-3}$ which is about half the mean of the $f_w$ as stated in the previous paragraph. 

Statistics of the duration at fixed frequency (Fig. \ref{distBW_Dur_Per} (c)) give an average value 0.28\,s, with very few values above 0.5\,s \citep[cf. ][for similar results]{Young61,Elgaroy73,Slottje1972,Bernold83}. The average repetition time (Fig. \ref{distBW_Dur_Per} (d)) is 0.90\,s, with few values above 1.5\,s \citep[see also][Figure 5]{Bernold83}.

Figure \ref{distBW_Dur_SingleFiber} shows distributions of the frequency range and total duration of fiber emission. In the frequency range histograms we have drawn separately regular (negative $\Delta f$) and reverse drift (positive $\Delta f$) bursts. The average value and the dispersion of the distributions are given in Table \ref{TableFibers}. The average value of the total duration is 3.7\,s with few fibers lasting more than 10\,s.

\section{Deduction of physical parameters from the observations} \label{Model}
\subsection{Estimate of the exciter speed and the magnetic field}\label{ExciterSpeed}
The exciter speed, $\rm{\upsilon}$, can be estimated from the frequency drift:
\be \label{Vexciter}
\upsilon  = \frac{{ds}}{{dt}} = \frac{{dz/\cos \theta }}{{dt}} =  - \frac{{{H_f}}}{{\cos \theta }}\frac{1}{f}\frac{{df}}{{dt}}
,\ee
where $H_f$ is the frequency scale height, $s$ is the distance along the path of the exciter, z is along the vertical and $\theta$ is the angle between the two. We note that $\upsilon$ in (\ref{Vexciter})  is independent of the nature of the exciter.

The frequency scale height is twice the density scale height and, for an isothermal corona:
\be \label{ScaleHeight01}
{H_f} = 2H = 2\frac{{kT}}{{\mu {m_p}{g_ \odot }}}{\left( {\frac{R}{{{R_ \odot }}}} \right)^2} \approx {\rm{100}}\, {\rm{T}}[\rm {MK}]\, {\left( {\frac{R}{{{R_ \odot }}}} \right)^2}\,\,\,[\rm {Mm}]
\ee
where $\mu$ is the mean molecular weight ($\approx 0.6$ in the corona), $k$ is the Boltzmann constant, $m_p$ is the proton mass, $T$ is the temperature, \GSUN ~is the acceleration of gravity in the photosphere and $R$ the distance of the emission region from the center of the sun ($R= z+R_ \odot$, with $z$ the height above the photosphere). For a given frequency, $R$ can be computed from a coronal density model.

\begin{figure}          
\begin{center}
\includegraphics[width=0.32\hsize]{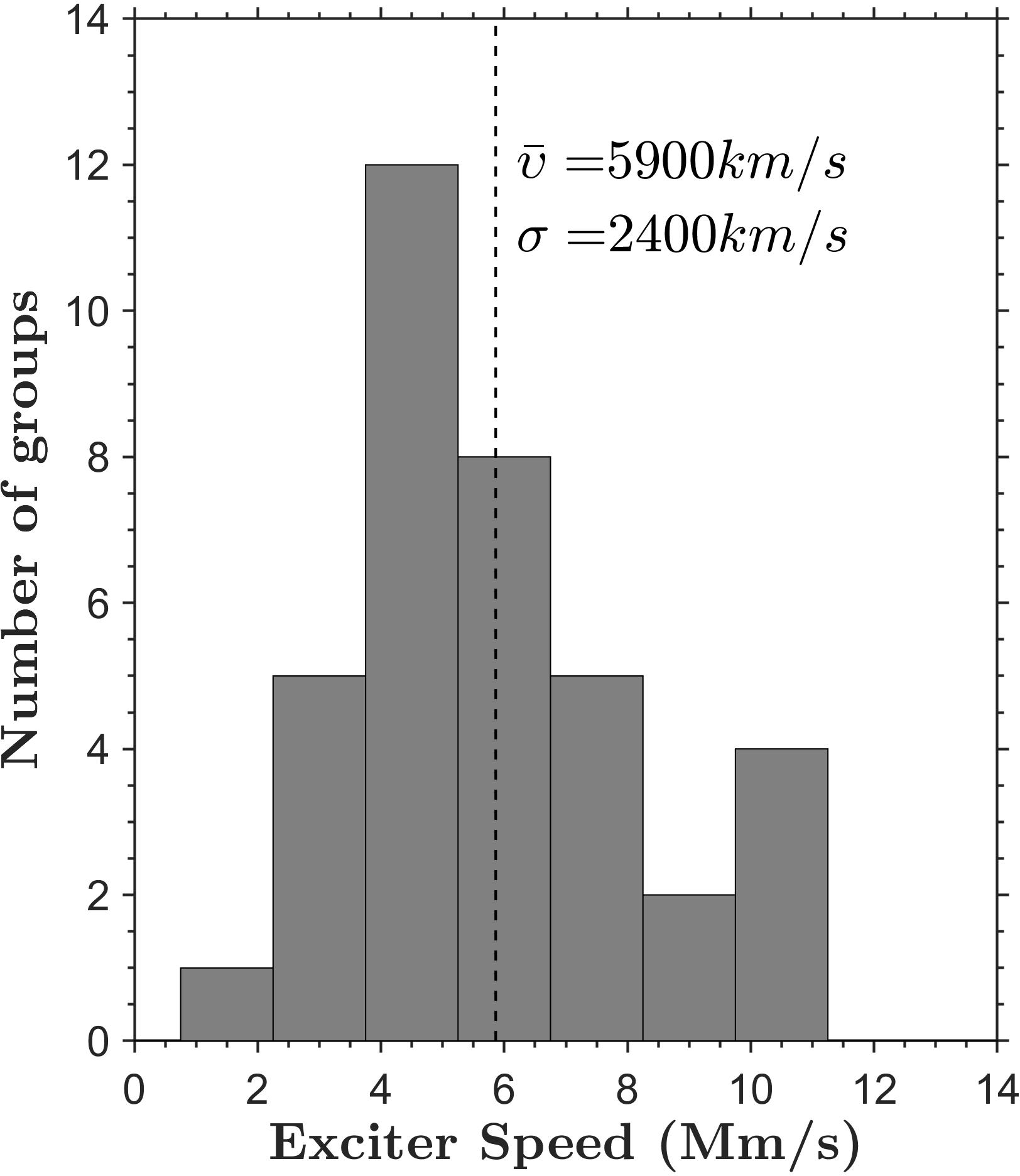}
\includegraphics[width=0.32\hsize]{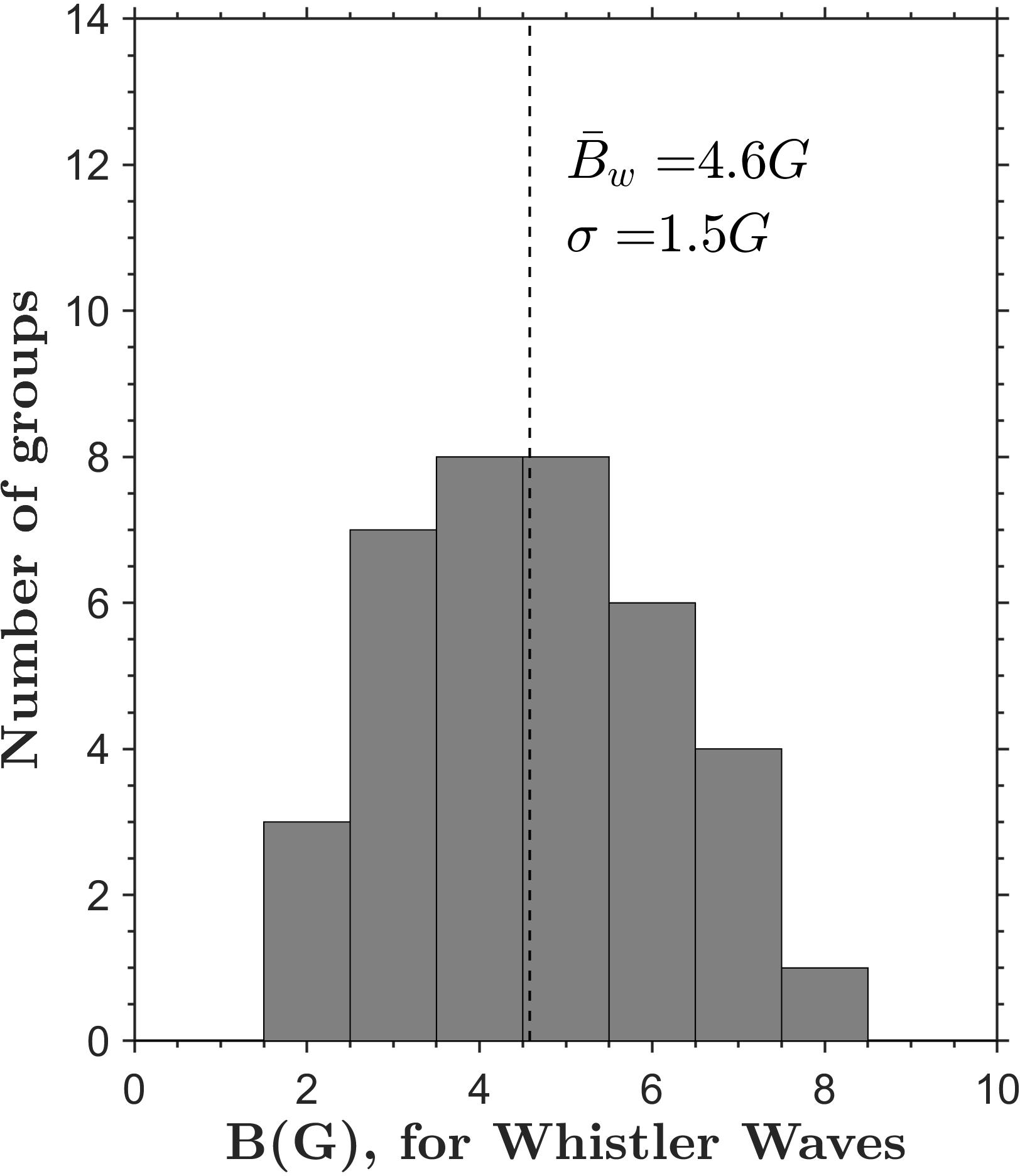}
\includegraphics[width=0.32\hsize]{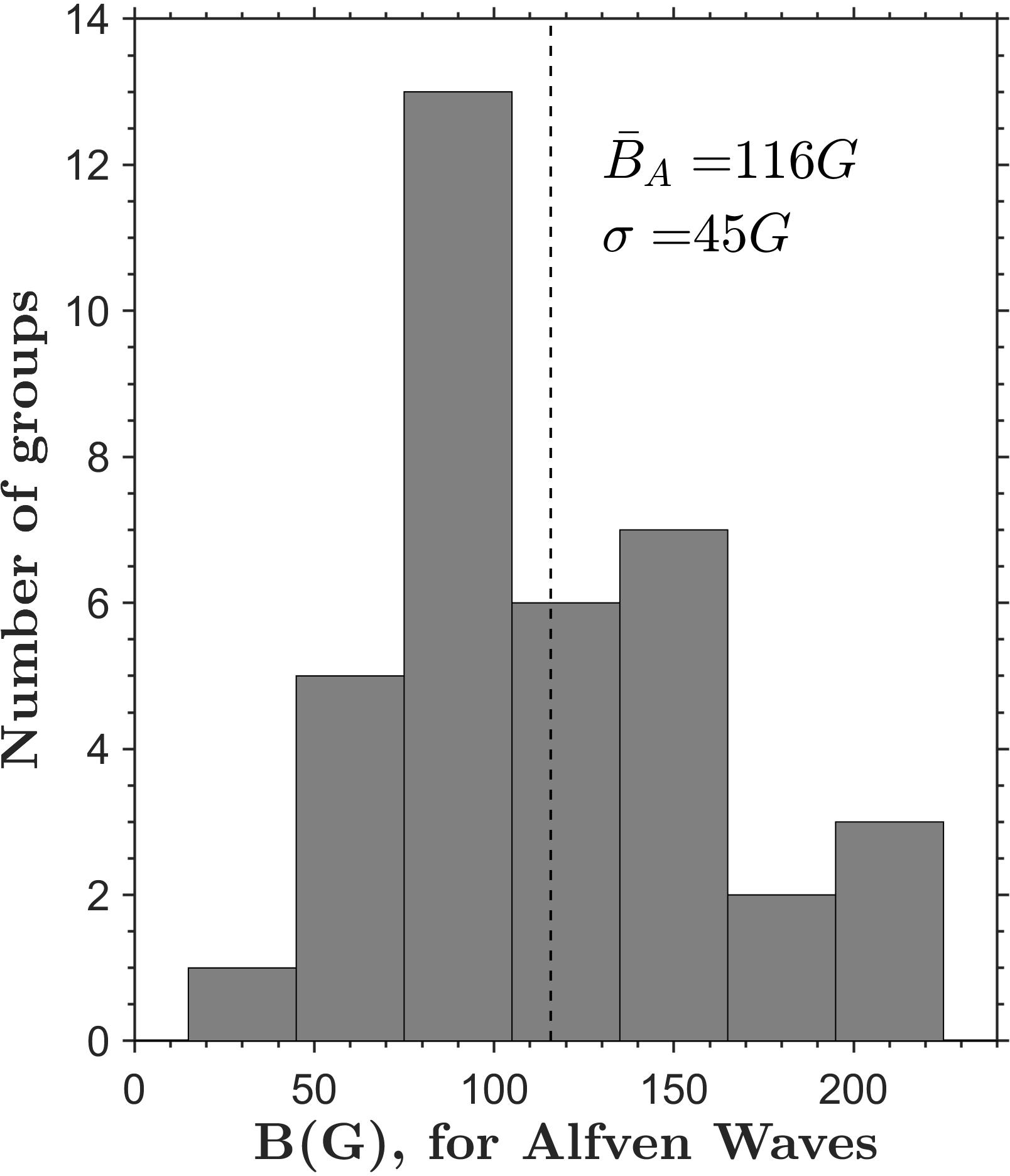}
\caption{Histograms of the exciter speed for \TypicalFibersNeg ~outbound typical fiber groups (left panel) and of the corresponding magnetic field assuming whistler~(middle panel) and \Alfven~waves (right panel)  as exciters.}
\label{SpeedHist}
\end{center}
\end{figure}

We used Eq. (\ref{Vexciter}) to estimate the exciter speed from the average frequency drift of the \TypicalFibersNeg ~typical fiber groups listed in Table \ref{Dataset}, measured by the 2D autocorrelation method described in Sect. \ref{DataAnalysis}. The frequency scale height was obtained from a hydrostatic density model with $T=2.0 \cdot 10^6 K$ and a base density four times that of the Newkirk model. The $\cos \theta$ factor was ignored, hence our values are lower limits of the emitter speed. We obtained an average value of \mbox{5900 km\,s$^{-1}$} with a dispersion of \mbox{2400 km\,s$^{-1}$} and the histogram shown in the left panel of Fig. \ref{SpeedHist}.

From the frequency drift we can estimate the magnetic field for specific types of exciters. Here we considered whistler-Langmuir wave-wave  interaction \citep{Kuijpers1972,Kuijpers1975,Kuijpers80}, and  \Alfven-Langmuir wave-wave interaction \citep{Bernold83}.

For whistler waves the group velocity is \citep{Kuijpers1975}:
\be\label{eq_whistler_velocity02}
v_{\rm{w}}  = 2c\left( {\frac{{\omega _{{\rm{ce}}} }}{{\omega _{{\rm{p}}} }}} \right)\sqrt {x\left( {1 - x} \right)^3 }  = 2c\left( {\frac{{f _{\rm{w}} }}{{f}}} \right)\sqrt {\frac{{\left( {1 - x} \right)^3 }}{x}}
\ee
where
\be
x =\frac {\omega_{\rm w}}{\omega_{\rm ce }}=\frac {f_{\rm w }}{2.8B}  \label{xdef}\,,
\ee
is the ratio of the whistler frequency to the electron-cyclotron frequency, where $f_{\rm w}$ is in MHz and $B$ in G. The whistler frequency can be measured from the frequency difference between absorption and emission ridges since the radiation is enhanced at $\omega_{\rm p}+\omega_{\rm w}$ and reduced at $\omega_{\rm p}$ due to the wave-wave interaction $l+w\rightarrow t$ (see Sect. \ref{Classification}). Substituting $v_w$ into Eq. (\ref{Vexciter}), we obtain for the drift rate:

\be\label{driftrate}
\frac{{df}}{{dt}} = -2cf_{\rm{w}} \frac{\cos\theta}{{{H_{\rm{f}}}}}\sqrt {\frac{{{{\left( {1 - x} \right)}^3}}}{x}}
.\ee
We note that, as the magnetic field decreases with height, $x$ will increase with time and the quantity under the square root in (\ref{driftrate}) will decrease in absolute value; on the other hand $H_f$ increases with height ({
cf.} Eq. \ref{ScaleHeight01}), $\cos\theta$ decreases, so that the effect of $\cos\theta/H_f$ will be an additional decrease of the absolute value of the drift with time, in agreement with the observations.

Estimating $H_f$ from a model, Eq. (\ref{driftrate}) can be solved for $x$ and hence for $B$; the result corresponds approximately to the average $B$ along the fiber. Ignoring again $\cos\theta$, we get a lower limit for $B$. For the same groups of fibers and the same density model, we obtained an average magnetic field of 4.6 G with dispersion of 1.5 G and the histogram shown in the middle panel of Fig. \ref{SpeedHist}.

Equating the exciter speed to the \Alfven~velocity,
\be \label{Alfven}
\upsilon _{\rm{A}}= \frac{B}{\sqrt{4\pi\rho}}
 {\rm{ = }}\frac{{\rm{B}}}{{\sqrt {{\rm{4\pi N}}_{\rm{e}} \mu {\rm{m}}_{\rm{p}} } }}
\ee
where $\rho$ is the plasma density and $N_e$ is the electron density, we can estimate the lower limit of the magnetic field under the assumption of \Alfven-Langmuir wave-wave interaction. The average value of the magnetic field is now 120 G with a dispersion of 50 G. The histogram is shown in the right panel of \mbox{Fig. \ref{SpeedHist}}.

The values of the magnetic field estimated above can be compared with measurements obtained from the circular polarization inversion in the microwave emission of active regions. According to the compilation of \citet[Table 1]{Alissandrakis1999} the magnetic field is $\approx$~10-15 G at a height of 100 Mm; also \citet{2017Carley} reported CME magnetic field of 4.4 G at a height of 1.3 \RSUN. The values of the magnetic field estimated above for \Alfven~waves are much higher than that, while those estimated from whistler waves are much closer. This result supports the whistler-Langmuir wave-wave interaction hypothesis as the source of the fiber radio emission; we also note that the study of fibers by \citet{2017Wang}~in the 1-2 GHz range precludes the hypothesis of \Alfven-Langmuir fiber origin. 

The above conclusion does not depend on the assumed parameters of the isothermal coronal model. As a test, we varied the base density between three and five times that of the Newkirk model and the ambient electron temperature in the  $T=1-3\cdot10^6$\,K range. The minimum magnetic field, under the assumption of \Alfven-Langmuir interaction, was obtained  for $T=10^6$\,K and a base density of 3xNewkirk model; the average value was 49 G with a dispersion of 19 G, which is still quite high. Thus, under the \Alfven-Langmuir hypothesis, a rather cool and underdense loop would be necessary to generate acceptable values of the coronal magnetic field.

\subsection{A simple geometrical model}\label{Geometrical}
Having selected the most likely mechanism, the track of a fiber on the dynamic spectrum (i.e., the frequency of the emission as a function of time) can be computed from the drift rate, provided that the path of the exciter in the corona and the local conditions (density and magnetic field) along the path are known.

Let us assume that the whistler waves propagate along a semicircular loop of radius $r$ \citep[cf.][]{Aschwanden2005}, so that
\be\label{cosTheta}
\cos \theta = \sqrt {1 - \frac{{{z^2}}}{{{r^2}}}}
\ee
and that the variation of the magnetic field along the loop is dipole-like:
\be\label{B_z}
B = {B_{\rm{0}}}{\left\{ {1 + \frac{z}{r}\left[ {\sqrt[3]{{\frac{{{B_0}}}{{{B_{\rm{t}}}}}}}{\rm{ - 1}}} \right]} \right\}^{{\rm{ - 3}}}} = {B_{\rm{0}}}{\left( {1 + Z\left( {\sqrt[3]{{{a_{\rm{m}}}}}{\rm{ - 1}}} \right)} \right)^{{\rm{ - 3}}}}
\ee
where $B_0$ is the magnetic field at the footpoints of the loop (\mbox{$z=0$}), $B_t$ the magnetic field at the top ($z=r$), $a_m={B_0}/{B_t}$ is the mirror ratio of the loop and $Z=z/r$ a dimensionless variable for the height. We add that the corresponding magnetic scale height, $H_{\rm m}$, is 
\be\label{H_m}
{H_m} \approx  - \frac{B}{{{{dB} \mathord{\left/
 {\vphantom {{dB} {dz}}} \right.
 \kern-\nulldelimiterspace} {dz}}}} = \frac{r}{3}\left( {\frac{{Z\left( {\sqrt[3]{{{a_{\rm{m}}}}} - 1} \right) + 1}}{{\sqrt[3]{{{a_{\rm{m}}}}} - 1}}} \right)
.\ee
\begin{figure}
\centering
\includegraphics[width=\hsize]{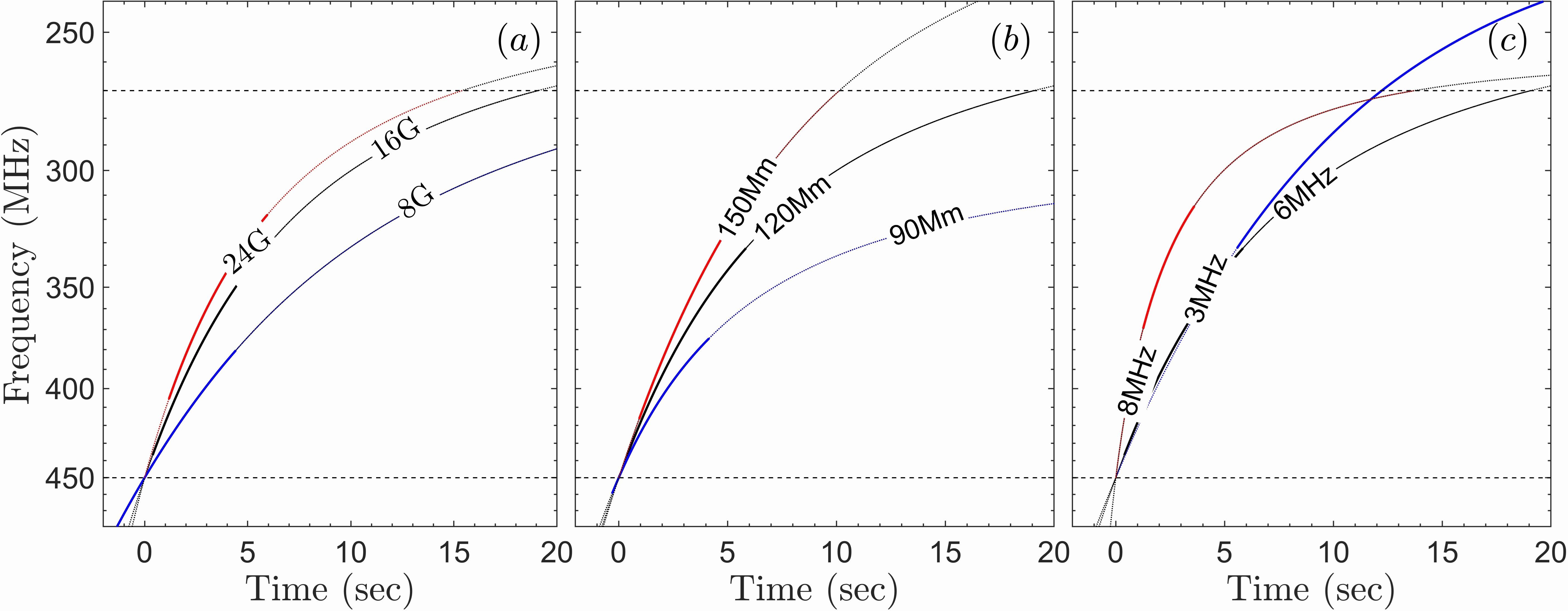}
\caption{Model tracks of fiber busts on dynamic spectra. The basic model (black line in all plots) is for $B_0=16$\,G, $r=120$\,Mm, $f_w=6$\,MHz, $B_t=2$\,G and a fourfold Newkirk density model. The left plot shows the effect of $B_0$, which is 8\,G for the blue line and 24\,G  for the red line, the middle plot shows the effect of the loop radius, which is 90\,Mm (blue line) and 150\,Mm (red line) an the right plot the effect of $f_w$, which is 3\,MHz for the blue line and 8\,MHz for the red. Thick lines denote the region for which $0.25<x<0.5$. Horizontal dashed lines mark the limits of the SAO frequency range.}
\label{FreqDriftMODEL}
\end{figure}

\begin{figure*}         
\centering                              
\includegraphics[width=.90\hsize]{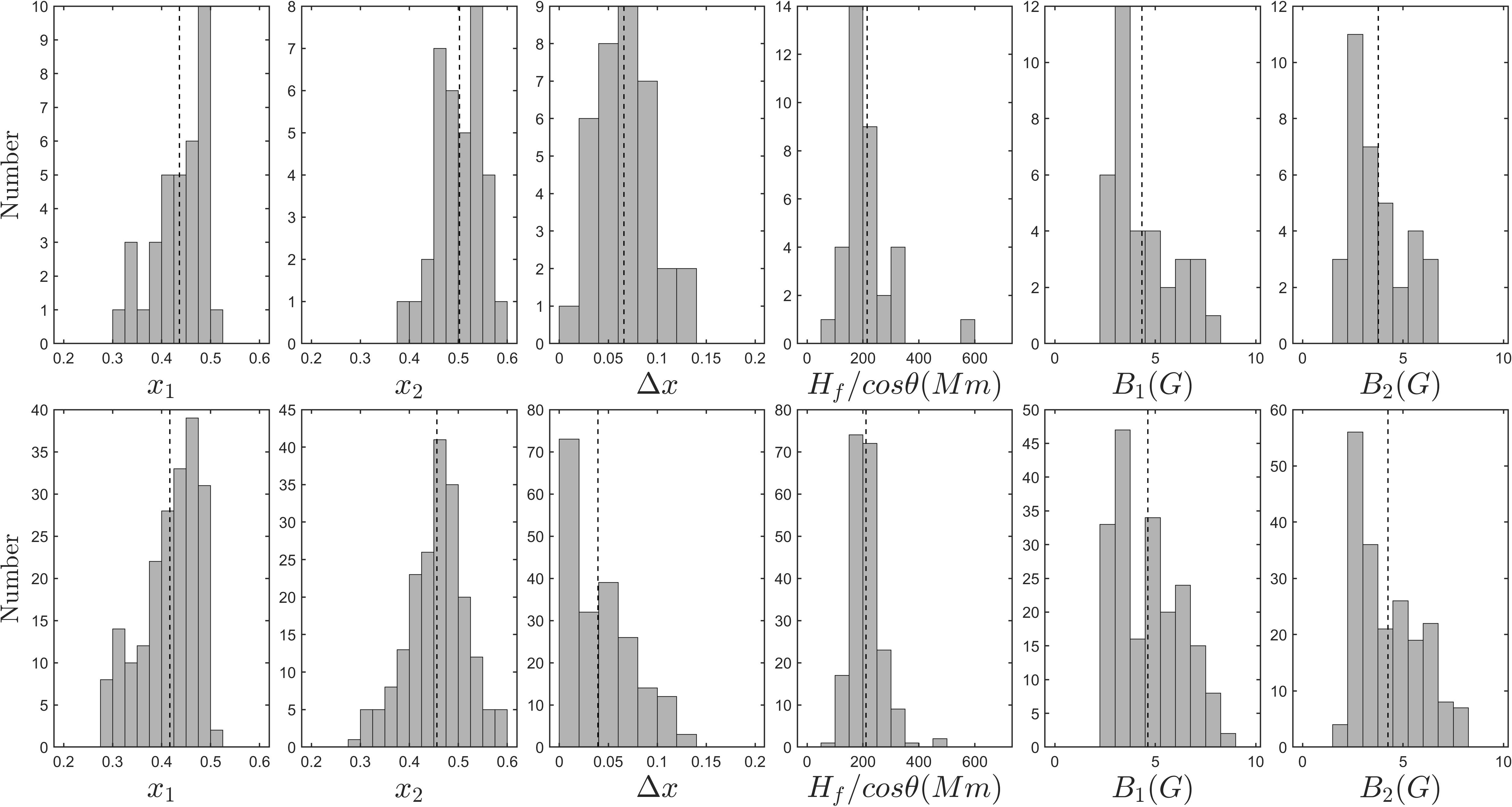}             
\caption{Histograms of fiber parameters derived from the least square fit of the drift rate of \TypicalFibersNeg ~outbound typical fiber groups (top) and the tracks of 209 individual fibers of the same class (bottom). $x_1,x_2$ are the start and end values of $x$, $\Delta x$ their difference, $H_f/\cos\theta$ the frequency scale length along the exciter trajectory, $B_1, B_2$ the magnetic field at the start and at the end of the fibers.}
\label{Hist2par}
\end{figure*}

Eqs.
(\ref{cosTheta}) and (\ref{B_z}), together with the value of the whistler frequency deduced from the observations, specify the values of $\theta$ and $s$ in Eq. (\ref{driftrate}) at height $z$. To compute the drift rate,
we further need to associate $z$ with a frequency and compute the corresponding value of $H_f$, and this requires the adoption of a coronal density model. The frequency as a function of time can be computed from the frequency drift by integration.

To illustrate the behavior of our model we present in Fig. \ref{FreqDriftMODEL} tracks of fibers on the dynamic spectrum for a basic model with $f_w=6$\,MHz, $B_0=16$\,G and $r=120$\,Mm, $B_t=2$\,G and the 4-fold Newkirk density model, as well as six variants to illustrate the effect of changing $B_0$, $r$ and $f_w$. The thick lines denote the region where  $0.25<x<0.5$. The frequency axis is plotted from high to low values so that the traces are easily comparable with observations.

The frequency-time curves resemble the observed dynamic spectra of fibers. Their inclination and curvature are affected by all three parameters, $B_0$, $r$ and $f_w$. Their duration  between 3 and 5\,s, is consistent with the \mbox{4.26$\pm$2.61 s} value in Table \ref{TableFibers} for the \TypicalFibersNeg ~groups of outbound typical fibers; the drift rate is between -30 and  -5 MHz\,s$^{-1}$, within the range of our measurements (Table \ref{Dataset}) for the same data set.

\subsection{Model-independent fiber parameters}\label{ModelInDependent}
Although the circular loop model predicts fiber-like tracks on the dynamic spectrum, it is not well adapted for a fit with the observations because it depends on the assumed models of the density and the magnetic field. Moreover, it has four free parameters (two for the magnetic field and two more for the density model); these are probably too many for a fit to the observed fiber tracks, which have a simple form that is well approximated by a second degree polynomial. 

Alternatively, the fiber shape in the dynamic spectrum can be modeled in terms of local values of the physical parameters involved. For example, our measurements of the drift rate as a function of time with the 1D cross-correlation method (Sect. \ref{DataAnalysis}) can be used to fit Eq. (\ref{driftrate}). The advantage of this procedure is that it is independent of any density and magnetic field model.

With $f_w$ specified as described above, Eq. (\ref{driftrate}) involves three fiber-related parameters: $\cos\theta$, $H_f$ and $x$. We first note that  $\cos\theta$ and $H_f$ cannot be determined independently from a least square fit, but their ratio can. It would be desirable to allow both $H_f/\cos\theta$ and $x$ to vary with time, but a simple linear variation of both requires four parameters for the fit; as discussed above, these are probably too many, considering the simple variation of the frequency drift with time. In the SAO frequency range (270\,-\,450 MHz) the frequency scale height calculated from the fourfold Newkirk model is in the range 190\,-\,150 Mm. As the total fiber bandwidth $\Delta f_{tot} \lesssim 55\,Mm$ for the outbound typical fibers (see Table \ref{TableFibers}), the relative variation of $H_f$ does not exceed 5\% and, if the whistler source is not too close to the loop-top, the variation of $\cos\theta$ is also expected to be quite small. The variation of the term $\sqrt {{{{{\left( {1 - x} \right)}^3}} \mathord{\left/{\vphantom {{{{\left( {1 - x} \right)}^3}} x}} \right. \kern-\nulldelimiterspace} x}}$, on the other hand, is expected to be $\gtrsim 50\%$, for $x$ varying in the 0.2\,-\,0.5 range. We may thus assume $H_f/\cos\theta$ constant and express x as a linear function of time:
\be
x(t)=x_1+\frac{x_2-x_1}{\Delta t} t = x_1+at
\ee
where $t$ is the time from the start of the fiber, $\Delta t$ its duration and $x_1, x_2$ the values of $x$ at the start and at the end. This gives us the drift rate in terms of three parameters, $H_f/\cos\theta$, $x_1$ and $a$, that can be determined from the observed drift through least square fit.

The same result can be {\bf derived}
with the measurements of the tracks of individual fibers, rather than the drift rate. Here things are more complicated, because we need to integrate Eq. (\ref{driftrate}) and this gives a rather cumbersome result, even with the assumption of constant $\cos\theta/H_f$:
\bea
f(t)= f_0 \!\!\!&-&\!\!\! \frac{c f_w}{2a} \frac{\cos \theta}{H_f} \left[ 3 \sin^{-1} (\sqrt{1-x_1}) + (2x_1-5) \sqrt{x_1(1-x_1)} \right. \nonumber \\
             \!\!\!&+&\!\!\! \frac{c f_w}{2a} \frac{\cos \theta}{H_f} \left[ 3 \sin^{-1} (\sqrt{1-x_1-at}) + (2at+2x_1-5) \right. \nonumber\\
             \!\!\!& \times &\!\!\! \left. \sqrt{ (at+x_1) (1-at-x_1)} \right]
.\eea

Histograms of the values of the derived parameters are given in Fig. \ref{Hist2par}. The two data sets (drift rate and fiber tracks) give very similar histograms, still the drift rate results should be more reliable because the associated measurements cover a wider frequency range. We first note that the average magnetic field (of about 4.44\,G) is very close to that derived in Sect. \ref{ExciterSpeed}. In addition, here we have an estimate of the range of $x$, which has a theoretical interest. The minimum derived value is 0.29, very close to the theoretical lower limit of 0.25 and the maximum value is 0.6, not much above the theoretical upper limit of 0.5; however, the range of $x$ in individual fibers or groups is rather small, not exceeding 0.12, which opens some questions as to what causes the start and the end of the whistler-Langmuir wave interaction that gives rise to the electromagnetic emission. 
It is possible that the range of $x$ represents an upper bound as Landau and cyclotron damping in the case of whistler propagation at an angle \citep[Sect. 3.3]{Kuijpers1975} is treated as negligible. Furthermore the effects of the dynamics of the energetic electrons trapped in the magnetic loop and the decrease of the loss-cone distribution with height might also result in a reduced range for $x$.

The average value of the deduced $H_f/\cos\theta$ $\sim220$\,Mm,  compared to average model value of 250 Mm used in Section \ref{ExciterSpeed}. It is greater than the Newkirk frequency scale height (190-150 Mm), since it was calculated along the non-radial exciter trajectory.
\begin{figure}
\centering
\includegraphics[width=0.95\hsize]{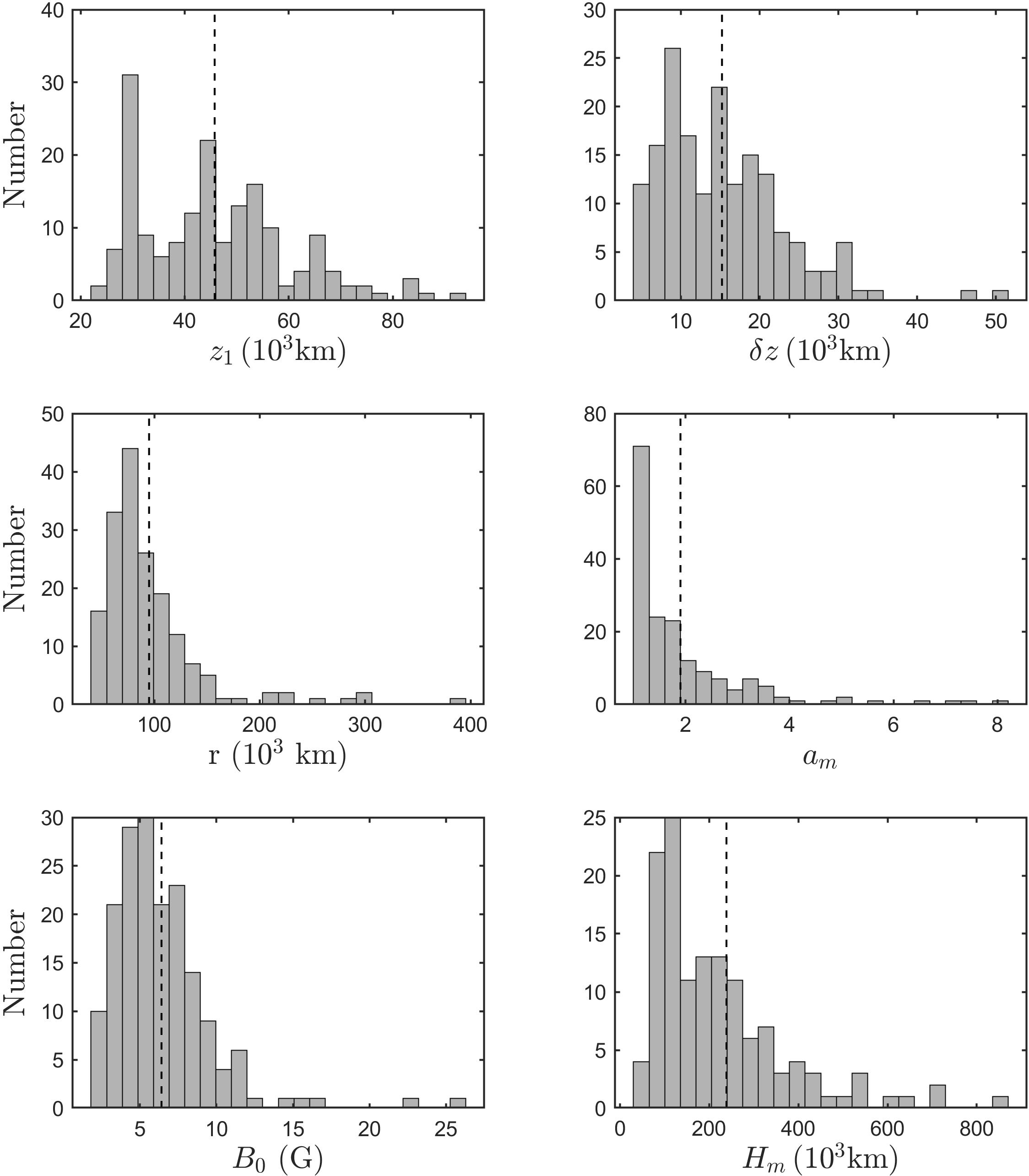}
\caption{Histograms of fiber parameters derived from the model dependent calculations of Sect. \ref{ModelDependent};~209 individual fibers were used. Top: Height, $z_1$, of the fiber onset and the extent in height, $\delta z$, from the fiber onset to the decay. Middle: Radius, $r$, of the assumed circular magnetic loop and the loop mirror ratio, $a_m$. Bottom: Magnetic field, $B_0$ at the loop footpoints and the magnetic scale height $H_m$. The dashed vertical lines in each histogram mark the median of the respective dataset.}
\label{HistModelDepend}
\end{figure}
\subsection{Model dependent fiber parameters}\label{ModelDependent}
Going one step further, from the {\bf frequency range} 
of the fiber emission and the adopted coronal density model we can derive the corresponding heights and from those the average frequency scale height. Combining this information  with the least square fit results for $H_f/\cos\theta$, we can estimate $\cos\theta$ and from that the semicircular loop radius, $r$, (Eq.\ref{cosTheta}). Then, Eq. \ref{B_z} and \ref{H_m} allow us to compute other parameters, such as the magnetic field at the loop footpoints, the magnetic scale height and the mirror ratio. 

Histograms of these parameters, derived from tracks of individual fibers, are given in Fig. \ref{HistModelDepend}. The fiber onset was found to be, on average, at a  height, \mbox{$z_1\approx46\,$Mm}, with an average vertical extent, \mbox{$\delta z\approx 15\,$Mm} The magnetic loop radius was found $r\approx50-200\,$Mm, with few cases of radii exceeding $250$\,Mm. The results are consistent with previous calculations by \citet{Aurass2005} and \citet{Rausche08}, of the 3D path of a few select fibers along magnetic lines. The average magnetic scale height (Eq. \ref{H_m}) and the loop mirror-ratio were found equal to 238\,Mm and 1.9 respectively.  This implies that the variation of the magnetic field with height is rather small, otherwise, given the restricted permissible range of $x$, the frequency extent of the fibers would have been significantly shorter.

\section{Relations between observed parameters}\label{relobs}
In this section we present empirical relations between measured parameters of intermediate drift bursts and their interepretation in terms of  the conditions of electron trapping and loss-cone distribution evolution.

\subsection{Repetition time and drift rate} \label{DriftPeriodicity}
{ Figure \ref{Relation_Per} shows a scatter plot of the repetition time of fibers
within an intermediate drift burst group, $T$, as a function of the average drift rate of the same group. The two quantities appear to be inversely proprotional to each other, so that:
\be \label{TDrift}
T \left| {\frac{{df}}{{fdt}}} \right|=\xi=0.0245 \pm 0.0025 
\ee
where the average value and the standard deviation of $\xi$ were computed from the values of Table \ref{Dataset} for typical fibers. When Eq. (\ref{TDrift}) was plotted in Fig. \ref{Relation_Per} it was found to extend to FDBs and ropes. 

Despite the large scatter in this relation, it is worth to examine if it has a theoretical justification. From Equations (\ref{eq_whistler_velocity02}) and (\ref{driftrate}) we have for the relative drift rate in the case of the semicircular loop model:
\be \label{RelDrift02}
\frac{{df}}{{fdt}} = -2c\frac{{\cos \theta }}{{H_f }}\frac{{f_w }}{f}\sqrt {\frac{{\left( {1 - x} \right)^3 }}{x}}  
.\ee
\begin{figure}[h]
\begin{center}
\includegraphics[trim=10.3cm 0cm  0cm 0.0cm,clip,width=1.10\hsize]{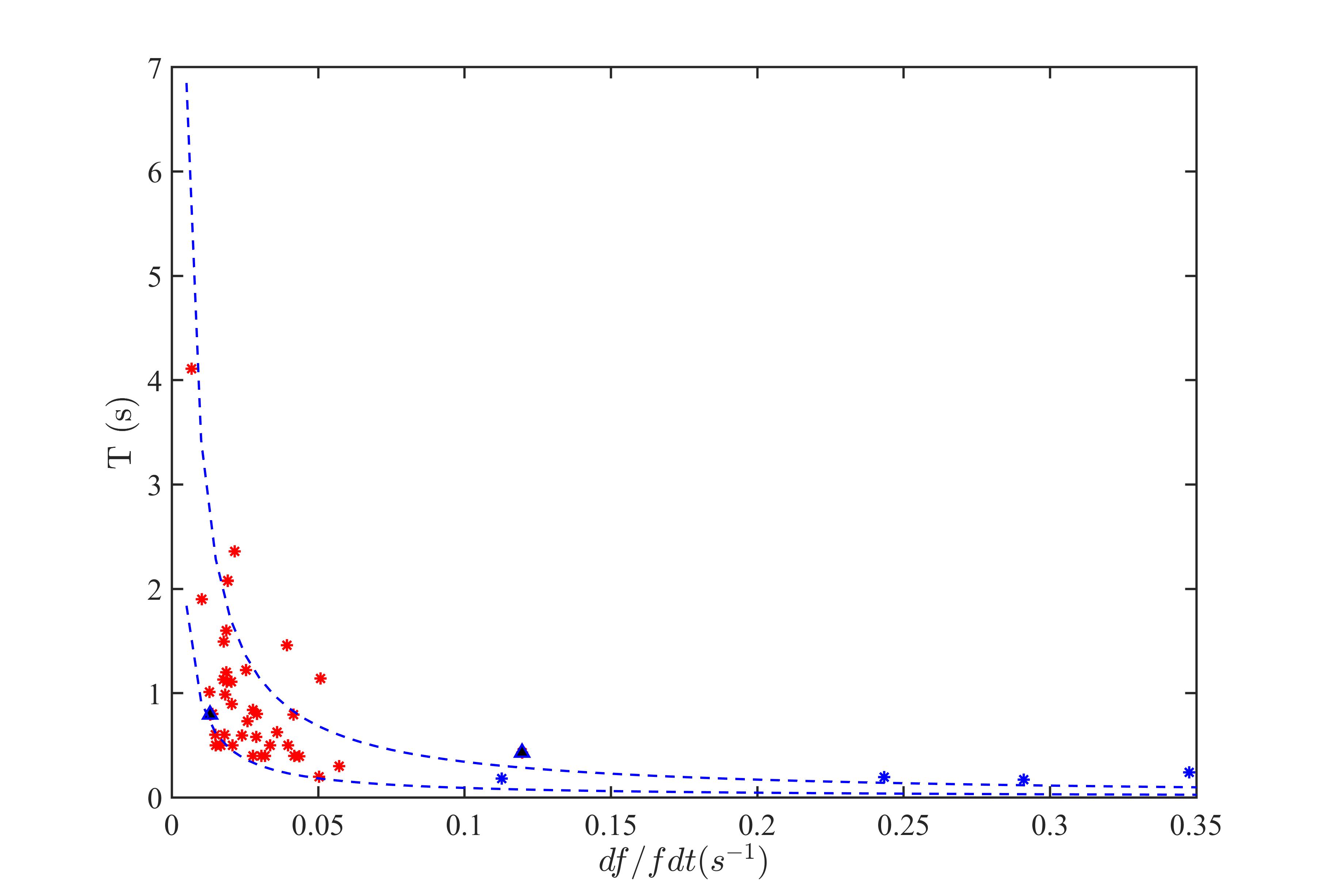} 
\caption{Fiber repetition time plotted against the relative drift rate. The blue lines represent the upper and lower bound of the dimensionless parameter, $\xi$, of Eq. \ref{TDrift}.  Typical fibers are shown as red stars, rope-like fibers as blue triangle and the FDBs as blue stars. } 
\label{Relation_Per}
\end{center}
\end{figure}
The fiber repetition time, based on the discussion in \citet[Sect. 4.2]{Kuijpers1975}, may be associated with the time required for the whistlers to move outside the initially unstable region of the loss cone and the latter to reform. Eq. \ref{TLC}, below provides an estimate of the time needed for the electrons with small pitch angle to escape the loop (expecting those with larger pitch angles to remain trapped) reforming a filled loss cone following the above mentioned relaxation of the distribution due to whistler-energetic electron interaction. 
\be
T \approx\ \tau _{LC} \approx r/v_\parallel
\label{TLC}
\ee
where $v_\parallel$ is the velocity of the energetic electrons parallel to the magnetic field. 

Based on the results of the model calculations in Sect. \ref{ModelDependent}, the parameters of fiber groups in Tables \ref{TableFibers} and \ref{Dataset} ($r \approx 100$\,Mm, $x=0.5$, $H_f/cos\theta\approx200$, \mbox{$f_w \approx 15 \cdot 10^{-3}\,f$)} and a typical value for the electron speed \citep[$\upsilon _\parallel \approx c/3$, see~][]{Kuijpers1975}, we can estimate the value of $\xi$:
\be \label{RelDrift03}
\xi = T\left| {\frac{{df}}{{fdt}}} \right| \approx 2c\frac{r}{{\upsilon _\parallel  }}\frac{{\cos \theta }}{{H_f }}\frac{{f_w }}{f}\sqrt {\frac{{\left( {1 - x} \right)^3 }}{x}}  \approx 0.0225
\ee
which is very close to the phenomenological value in Eq. \ref{TDrift}.
 \begin{figure}
 \begin{center}
 \includegraphics[trim=7.0cm 0cm  0cm 0.0cm,clip,width=1.10\hsize]{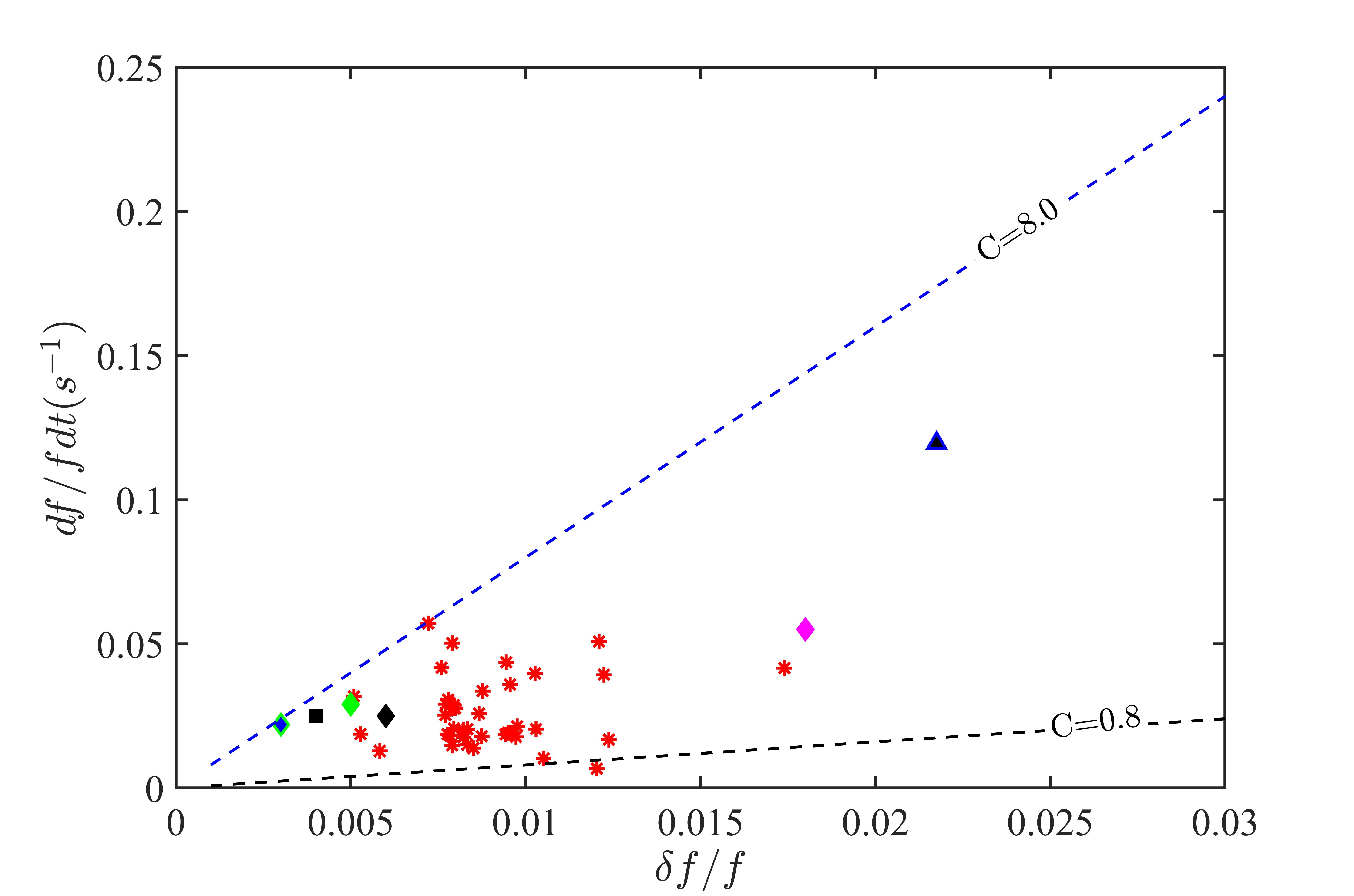} 
 \caption{Relative drift rate ($df/fdt$) versus relative instantaneous bandwidth ($\delta f/f$). The data points corresponding to typical fiber groups are the red stars and the single rope-like group is the blue triangle. The  fiber burst  points from \citet{Young61} (magenta diamond), \citet{Elgaroy73} (blue and green diamond), \citet{Slottje1972, Bernold83} (black diamond) are included in the same graph for comparison. The rope-like fibers from \citet{Chernov2008} are indicated with the black square. The dashed lines are from Eq. \ref{RelDrift04} with $C=0.8s^{-1}$  and $C=8.0s^{-1}$.} 
 \label{RelBW_Drift}
 \end{center}
 \end{figure}
 \begin{figure}
 \begin{center}
 \includegraphics[trim=8.0cm 0cm  0cm 0.0cm,clip,width=1.10\hsize]{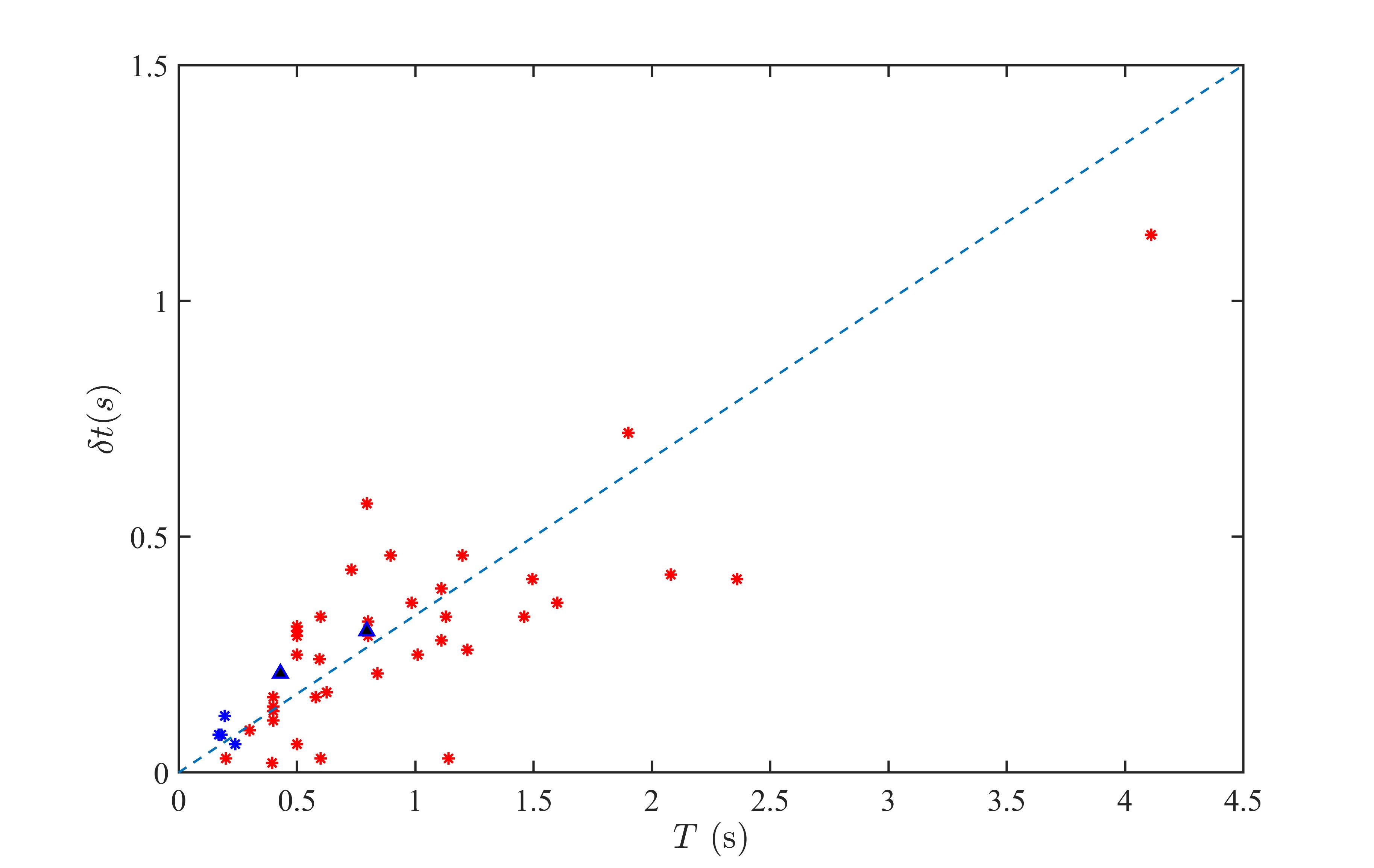} 
 \caption{Duration at fixed frequency ($\delta t$) versus repetition time (T). The blue dashed line represents the relationship from Eq. \ref{DtT}. Rope-like fibers (blue triangle) and fast drift bursts (blue stars) are included. }
 \label{Duration_Per}
 \end{center}
 \end{figure}
\subsection{Fiber drift rate and instantaneous bandwidth} \label{DriftBW}
{The relative instantaneous bandwidth ($\delta f/f$) is plotted versus the absolute value of relative drift rate ($\delta f\delta t/f$) in Fig. \ref{RelBW_Drift}. Data points from \citet{Young61}, \citet{Elgaroy73}, \citet{Slottje1972, Bernold83} and \citet{Chernov2008} included for comparison.

 \citet[(see their Eq. 2)]{ElgaroySoldal1981} were the first to obtain a linear fit on this type of observation in the 310-340 MHz frequency range:}
 {\be\label{ElgaroySoldal1981Eq}
\Delta f\left( {{\rm{MHz}}} \right) = 0.5\left( {{\rm{MHz}}} \right) - 0.2\left( {\rm{s}} \right)\left( {\frac{{df\left( {{\rm{MHz}}} \right)}}{{dt\left( {\rm{s}} \right)}}} \right)
 .\ee}
Hence:
 \be\label{ElgaroySoldal1981EqB}
\frac{{df}}{{dt}} = 5\left( {s^{{\rm{ - 1}}} } \right)\left( {\delta f - 0.5} \right)
 .\ee
 
 For comparison with (\ref{ElgaroySoldal1981Eq}) and (\ref{ElgaroySoldal1981EqB}) and  using $f_w$ instead of $\delta f$,  we may rewrite Eq.(\ref{RelDrift02}) in the form:
 \be\label{RelDrift04}
\left| {\frac{{df}}{{fdt}}} \right| = \left( {2c\frac{{\cos \theta }}{{H_f }}\sqrt {\frac{{\left( {1 - x} \right)^3 }}{x}} } \right) \cdot \left( {\frac{{f_w }}{f}} \right) = C \cdot \frac{{\delta f}}{f}
 \ee
where, in the rightmost substitution, we took into account that $f_w\approx 2\cdot \delta f$, because the instantaneous bandwidth ($\delta f$) has been measured as the full width at half maximum of the emission ridge and the whistler frequency ($f_w$) as the frequency difference between the absorption and emission peaks (Sect \ref{SingleBurst}).

We estimated upper and lower bounds of the constant C which are shown in Fig. \ref{RelBW_Drift}. In this case the value of the bounds was calculated from eq. \ref{RelDrift04}, with average values characteristic of typical fiber groups in Table \ref{Dataset} (\mbox{{$100$~(Mm)$ \le H_f/cos\theta\le300~$(Mm)}}, \mbox{$0.20\le\sqrt {{\left( {1 - x} \right)^3 }/{x}}\le0.73$, $0.30\le x\le0.70$}). The bounds were computed to \mbox{$0.80\le C\le8.0$}; the results are broadly consistent with the empirical relationship of  \citet[Eq. 2]{ElgaroySoldal1981}.

The range of the constant C, corresponding to the scatter of the  relative instantaneous bandwidth values on Fig. \ref{RelBW_Drift}, might be due to the 1.4 MHz frequency width of our SAO receiver which is close to the measured instantaneous bandwidth $\delta f$ with mean value equal to 2.4 MHz for typical fibers.

\subsection{Repetition time and duration at fixed frequency}
{The duration at fixed frequency, $\delta t$, of individual fiber events within a group is directly linked to the instantaneous bandwidth, $\delta f$, and the frequency drift rate: 
\be     
        \delta t = \frac{{\delta f}}{f}{\left| {\frac{{df}}{{fdt}}} \right|^{ - 1}}
.\ee

The repetition time, $T$, within a group is linked to the relative drift rate through Eq. \ref{RelDrift03}. Hence we have
\be \label{DtT}
\delta t \approx \frac{1}{2}\left( {\frac{{f_w }}{f}} \right)\frac{T}{{\xi }} = 0.33\, T
\ee
where we have set \mbox{$f_w/f\approx 15\cdot10^{-3}$} from Table \ref{TableFibers}.

Figure \ref{Duration_Per} presents the scatter plot of measured  duration at fixed frequency versus the repetition time, which is consistent with the linear relationship of Eq. \ref{DtT}; the rope-like fibers and FDBs, included in this plot, are not separate from the typical fibers. 
}
 
\section{Summary and conclusions} \label{discussion}
In this work we have used high time resolution dynamic spectra for the study of intermediate drift bursts, which constitute a significant component of the fine structure of type IV radio bursts. The typical fiber bursts, which comprise the vast majority of this type of fine structure,  were divided in six morphological groups, based on the position of the emission and the absorption ridges. Among these groups a rather rare type of fibers with absorption ridge at the higher frequency side of the dynamic spectrum and emission ridge at the lower frequency was found; this is the opposite of the expected absorption-emission order in frequency.  Another rare morphological type with an absorption ridge between two emission ridges was also recorded. In addition to the typical fiber bursts, a small number of narrowband fibers, rope-like fibers and fast-drift burst groups were observed. These were also characterized by an emission-absorption structure similar to the typical fibers.

A 2D autocorrelation analysis of the dynamic spectra of fiber burst groups was employed for the measurement of the average frequency drift rate, the instantaneous bandwidth and the duration at fixed frequency. The same type of analysis gave the average time interval between successive fibers in the group.  The variation of the drift rate with frequency was measured by means of a window, sliding along the frequency axis of the spectra, and the computation of the cross correlation of each frequency channel with respect to the central one. {Individual fibers were examined for the determination of the total frequency extent and duration of fibers within groups. The results are \mbox{given in Tables \ref{TableFibers}}  and \ref{Dataset}.

We developed a semi-automatic algorithm to track individual fibers and applied it to a sample of $\sim200$ events. This gave results similar to those of the 2D autocorrelation method. }
 
Based on our measurements and a four-fold Newkirk model, we estimate an average exciter speed of $5900 \pm 2400$ km\,s$^{-1}$; this gives an average magnetic field of $4.6 \pm 1.5$\,G for whistler-driven fibers and $116 \pm 50$\,G for Alfven-wave-driven fibers; the former being much closer to values estimated by other methods, we consider the whister mechanism as the most plausible one.

We find that a simple semi-circular loop model with dipole-like variation of the magnetic field with height reproduced the shape of fibers on the dynamic spectra. However, this model required more free parameters than the ones that could be determined from the observations. Therefore, we resorted to a simpler model that allowed the magnetic field to vary along the exciter trajectory, without assuming any particular loop geometry and we determined the frequency scale height along the path ($\sim 220$\,Mm on average) and the value of the ratio of the whistler frequency to the electron cyclotron frequency, $x$, which is in the range of 0.29 to 0.6.

For a fourfold Newkirk model and the frequency range of fibers, we estimated that their onset is at a height of  $\sim 46$\,Mm and their vertical extent is $\sim 15$\,Mm. Feeding these results to the circular loop model, we obtained a magnetic field scale height of $\sim 240$\,Mm and mirror ratio of  $\sim 1.9$.
  
Empirical relationships between observational characteristics were examined: The repetition time and the relative drift rate were found to be inversely proportional, with considerable scatter however. The plot of the relative drift rate 
versus the instantaneous banwidth was limited by two staight lines. Finally, the fiber duration at fixed frequency showed a well defined association with the repetetion time. We interpret these relationships in terms of the semicircular loop model.
 
The observational results of this work confirm the long-standing ideas on the origin of fiber bursts, characterized by a combination of absorption and emission ridges of almost equal instantaneous bandwidth, from three wave interactions (Langmuir (l), whistler (w), and transverse (t) wave) of the \mbox{$\rm l+w\rightarrow t$} and  \mbox{$\rm l\rightarrow t+w$} type \citep{Kuijpers1975, Chernov2011}. Outside these categories, a number of complex groups of overlapping intermediate drift bursts, mostly mixed groups of outbound and inbound fibers, at drift rates varying within each group was detected; the relationship of the various groups sources is not clear at this point and a circumstantial overlay on the dynamic spectrum cannot be excluded. Moreover, our observations provided valuable diagnostics of the low corona, the magnetic field in particular, which is inaccessible from other spectral regions apart from the radio band. 

Further research into the issue requires combination of  high--resolution and high--cadence radio spectra with simultaneous radio images.This will be the subject of a follow-up article, in which we will use images from the Nan\c cay Radioheliograph in conjunction with ARTEMIS/JLS spectral data.
\begin{acknowledgements} 
We wish to thank the anonymous reviewer for his/her careful reading of the article and his/her suggestions, which have improved the quality of this article. 
\end{acknowledgements}
\bibliographystyle{aa}
\bibliography{Ref/P01,Ref/P02,Ref/P03,Ref/P04,Ref/P05,Ref/P06,Ref/P07,Ref/P08,Ref/BurstsII,Ref/BurstsIII,Ref/BurstsIV,Ref/General,Ref/LNP725,Ref/SEE2007,Ref/Books,Ref/Density}

\begin{appendix}
\section{Parameters of intermediate drift bursts groups}\label{Appendix01}
In Table \ref{Dataset} we give a summary of the characteristics of the \TotalEvents ~intermediate drift burst groups observed with the \mbox{Artemis-JLS/SAO} receiver in the 270-450 MHz range.
{ The first columns  of the table present observational characteristics of each group:} Columns 2 and 3 give the date and start time. Column 4 gives the fiber duration at fixed frequency ($\delta t$), averaged over all fibers of each group. The fiber repetition rate (T) is given in column 5. An estimate of the whistler frequency ($f_w$) calculated from the frequency difference between absorption and emission peaks is given in column 6 with the whistler frequency normalized to the average of the observation frequency ($f_w/f$) in column 7. The frequency drift rate ($df/dt$) of the fibers and the corresponding normalized drift rate ($df/fdt$) are in columns 8 and 9.

The final columns contain parameters derived from least square fit of the model  (Sect. \ref{Model}) on the recorded fiber time-frequency tracks of the \TypicalFibersNeg ~typical fiber groups with negative frequency drift: Columns 10-12  give the magnetic field at the starting point of the individual fiber bursts, $B_1$, the ratio of the whistler to the cyclotron frequency, $x_1$, at the same point and the frequency scale length along the exciter trajectory, $H_f/\cos\theta$; the vast majority is of the class with an emission ridge and a lower frequency (LF) absorption ridge as noted in Sect. \ref{TypicalFibers}.

\begin{table*}[!htb]
\centering
\caption{Measured Characteristics of each Intermediate Drift Burst Group}
\label{Dataset}
\begin{tabular}{lll    SSS    SSS  SSS}          
\hline
\#	&\CC{Date}			&\CC{Start}		&\CC{$\delta t$} &\CC{$T$}	&\CC{$f_w$}	&\CC{${f_w/f}$}  	&\CC{$df/dt$} 				&\CC{$df/fdt$}				&\CC{$B_1$} & \CC{$x_1$} 	& \CC{$H_f /\cos\theta$}  \\ %
	&					&\CC{UT}		&\CC{(s)}		 &\CC{(s)}	&\CC{MHz}	&\CC{($10^{-3}$)} 	&\CC{MHz$ \cdot s^{-1}$}	&\CC{$10^{-2}s^{-1}$}		&\CC{(G)}	& 				&	\CC{Mm} \\
\hline
\hline
\multicolumn{12}{c}{\emph{Typical Fibers}}\\
\hline
01	&	{30/04/2000}	&	{07:58:18}	&	0.41		&	2.36	&	6.0		&	14.54			&	-08.91			&	-2.14	&	4.7	&	0.45&	219 	\\
02	&	{		}		&	{08:13:04}	&	0.17		&	0.63	&	3.0 	&	09.98 			&	-11.40			&	-3.59	&	3.7	&	0.30&	158 	\\
03	&	{11/07/2000}	&	{13:20:10}	&	0.41		&	1.50	&	3.5 	&	11.37			&	-05.57			&	-1.77	&	2.8	&	0.45&	215 	\\
04	&	{		}		&	{13:20:17}	&	0.36		&	1.60	&	3.8		&	11.82			&	-05.93			&	-1.86	&	3.1	&	0.44&	213 	\\
05	&	{		}		&	{13:20:40}	&	0.42		&	2.08	&	5.0		&	15.83			&	-06.08			&	-1.92	&	3.8	&	0.47&	244 	\\
06	&	{		}		&	{13:32:32}	&	0.25		&	1.01	&	3.1		&	09.02 			&	-04.85			&	-1.29	&	2.4	&	0.47&	236 	\\
07	&	{		}		&	{13:34:35}	&	0.33		&	1.13	&	3.3		&	09.16 			&	-06.74			&	-1.76	&	2.9	&	0.41&	191 	\\
08	&	{		}		&	{13:35:40}	&	0.28		&	1.11	&	3.3		&	08.65 			&	-07.21			&	-1.87	&	3.0	&	0.39&	189 	\\
09	&	{		}		&	{13:52:55}	&	0.26		&	1.22	&	6.1		&	15.73			&	-09.88			&	-2.53	&	4.9	&	0.44&	211 	\\
10	&	{14/07/2000}	&	{10:36:30}	&	0.43		&	0.73	&	8.6		&	24.84			&	-08.92			&	-2.58	&	6.3	&	0.49&	277 	\\
11	&	{		}		&	{10:39:37}	&	0.33		&	1.46	&	6.3		&	19.15			&	-12.80			&	-3.93	&	5.6	&	0.40&	188		\\
12  &	{		}		&	{{11:00:38}}	&0.25		&	0.61	&	6.3		&	17.40			&	-09.33			&	-2.13	&	5.0	&	0.45&	248 	\\
13	&	{18/11/2000}	&	{13:36:17}	&	0.14		&	1.18	&	8.5		&	25.28			&	-17.50			&	-5.08	&	6.9	&	0.44&	213 	\\
14	&	{21/04/2003}	&	{13:06:26}	&	0.11		&	0.30	&	4.5		&	10.34			&	-24.10			&	-5.71	&	5.2	&	0.31&	124 	\\
15	&	{		}		&	{13:10:07}	&	0.09		&	0.68	&	5.1		&	16.17			&	-13.90			&	-4.36	&	5.2	&	0.36&	178 	\\
16	&	{28/10/2003}	&	{11:13:59}	&	0.16		&	0.58	&	3.3		&	08.71 			&	-10.90			&	-2.88	&	3.7	&	0.32&	150 	\\
17	&	{		}		&	{11:17:15}	&	0.21		&	0.84	&	3.5		&	09.36 			&	-10.40			&	-2.76	&	3.8	&	0.33&	157 	\\
18	&	{		}		&	{12:00:07}	&	0.25		&	0.50	&	5.8		&	17.92			&	-05.37			&	-1.68	&	4.4	&	0.47&	285 	\\
19	&	{04/02/2004}	&	{11:18:52}	&	0.20		&	0.80	&	7.8		&	19.64			&	-16.60			&	-4.18	&	6.9	&	0.40&	195 	\\
20	&	{		}		&	{11:19:55}	&	0.10		&	0.41	&	8.4		&	22.20			&	-19.10			&	-5.02	&	8.3	&	0.36&	180 	\\
21	&	{06/04/2004}	&	{13:24:48}	&	0.17		&	1.21	&	3.6		&	10.59			&	-06.15			&	-1.79	&	2.9	&	0.44&	217 	\\
22	&	{13/07/2004}	&	{09:04:25}	&	0.06		&	0.50	&	8.7		&	25.34			&	-11.80			&	-3.36	&	6.5	&	0.47&	250 	\\
23	&	{		}		&	{09:27:35}	&	0.29		&	0.50	&	8.9		&	23.51			&	-07.85			&	-2.07	&	6.4	&	0.49&	307 	\\
24	&	{		}		&	{09:28:16}	&	0.46		&	1.20	&	4.4		&	11.42			&	-07.23			&	-1.86	&	3.5	&	0.44&	215 	\\
25	&	{15/01/2005}	&	{07:58:35}	&	0.36		&	0.99	&	4.5		&	11.85			&	-06.94			&	-1.82	&	3.7	&	0.44&	221 	\\
26	&	{17/01/2005}	&	{09:22:33}	&	0.19		&	0.50	&	8.8		&	22.69			&	-15.70			&	-3.97	&	7.4	&	0.43&	202 	\\
27	&	{		}		&	{09:30:11}	&	0.17		&	0.40	&	8.9		&	23.66			&	-10.80			&	-2.76	&	6.7	&	0.47&	249 	\\
28	&	{		}		&	{09:30:23}	&	0.46		&	0.90	&	7.5		&	19.25			&	-08.09			&	-2.05	&	5.5	&	0.49&	281 	\\
29	&	{		}		&	{09:31:54}	&	0.30		&	0.50	&	6.1		&	16.92			&	-05.33			&	-1.50	&	4.4	&	0.50&	373 	\\
30	&	{		}		&	{09:39:32}	&	0.32		&	0.80	&	5.0		&	12.89			&	-11.40			&	-2.91	&	4.7	&	0.39&	181 	\\
31	&	{		}		&	{09:45:12}	&	0.39		&	1.11	&	3.5		&	09.78 			&	-07.27			&	-2.04	&	3.2	&	0.40&	184 	\\
32	&	{		}		&	{09:45:45}	&	0.72		&	1.90	&	3.1		&	09.57 			&	-02.89			&	-1.03	&	3.0	&	0.37&	180 	\\
33	&	{20/01/2005}	&	{06:55:30}	&	1.14		&	4.11	&	4.9		&	14.74			&	-02.26			&	-0.68	&	3.5	&	0.50&	450 	\\
34	&	{13/07/2005}	&	{14:04:18}	&	0.38		&	0.80	&	9.5		&	27.62			&	-15.60			&	-4.16	&	7.4	&	0.46&	221 	\\
35	&	{14/07/2005}	&	{11:26:08}	&	0.13		&	0.40	&	3.1		&	07.99 			&	-12.50			&	-3.18	&	3.7	&	0.30&	136 	\\
36	&	{		}		&	{11:28:56}	&	0.33		&	0.60	&	3.4		&	08.83 			&	-05.61			&	-1.48	&	2.7	&	0.45&	218 	\\
37	&	{		}		&	{11:38:59}	&	0.16		&	0.40	&	5.7		&	14.86			&	-12.30			&	-3.06	&	5.1	&	0.40&	191 	\\
38	&	{		}		&	{11:52:24}	&	0.29		&	0.80	&	4.1		&	11.62			&	-04.85			&	-1.39	&	3.1	&	0.48&	266 	\\
39	&	{01/08/2010}	&	{08:18:10}	&	0.24		&	0.60	&	6.4		&	17.57			&	 08.82			&	 2.39	&	 	&		&			\\
\hline                                                                                               
\hline
\multicolumn{12}{c}{\emph{Rope-like Fiber Bursts}}		\\
\hline
40	&	{15/04/2000}	&	{10:17:03}	&	0.21		&	0.43	&	6.90	&	25.10 			&	-33.0			&	-12.0	&	 	&		&			\\
\hline
\hline
\multicolumn{12}{c}{\emph{Fast Drift Fiber Bursts}}	\\
\hline
41	&	{11/07/2000}	&	{13:18:26}	&	0.06		&	0.24	&	7.50	&	24.70			&	-108.0			&	-34.8 	&		&		&			\\
42	&	{		}		&	{13:21:41}	&	0.08		&	0.18	&	4.91	&	13.70			&	42.6			&	11.3  	&		&		&			\\
43	&	{		}		&	{13:52:11}	&	0.08		&	0.17	&	6.95	&	21.10			&	95.7			&	29.1  	&		&		&			\\
44	&	{17/01/2005}	&	{09:20:40}	&	0.12		&	0.20	&	8.01	&	22.20			&	82.4			&	24.3  	&		&		&			\\
\hline
\hline

\multicolumn{12}{c}{\emph{Narrow-band Intermediate Drift Bursts}}		\\
\hline
45	&	{11/07/2000}	&	{13:23:10}	&	0.30		&	0.84	&	3.18	&	09.35			& 	-5.86			&	-1.42 	&		&		&			\\
46	&	{21/04/2003}	&	{13:07:40}	&	0.30		&	0.80	&	4.50	&	12.50			&	4.78 			&	1.30  	&		&		&			\\
47	&	{14/07/2005}	&	{11:25:21}	&	0.85		&\CC{-}		&	2.16	&	05.68			& 	1.84			&	0.44  	&		&		&			\\
                                                                                    				
\hline
\end{tabular}
\end{table*}

\end{appendix}

\end{document}